\documentclass[aps, prl, twocolumn, superscriptaddress]{revtex4-1} 
\usepackage{amssymb, amsmath, amsthm}
\usepackage{xcolor}
\usepackage{graphicx}
\usepackage{hyperref}
\usepackage{comment}
\usepackage{blkarray}
\usepackage{mathtools}
\usepackage{tikz}
\usetikzlibrary{decorations.pathreplacing}
\usepackage{dsfont}
\usepackage{braket}
\usepackage[caption=false]{subfig}
\usepackage{float}
\usetikzlibrary{decorations.pathreplacing}
\usepackage{dsfont}
\usepackage{braket}

\usepackage{graphicx}
\usepackage{dcolumn}
\usepackage{bm}

\newcommand{\prlsection}[1]{{\em {#1}.---~}}
\newcommand{\Tr}[1]{\mathrm{Tr}}

\renewcommand{\vec}[1]{\boldsymbol #1}

\usepackage{bm}
\newcommand{\be}{\begin{equation}}
\newcommand{\ee}{\end{equation}}

\newcommand{\titleinfo}{
Three-stage thermalisation of a quasi-integrable system }
\begin{document}
\preprint{APS/123-QED}
\title{\titleinfo}

\author{Leonardo Biagetti}

\author{Guillaume Cecile}

\author{Jacopo De Nardis}
\affiliation{Laboratoire de Physique Th\'eorique et Mod\'elisation, CNRS UMR 8089,
	CY Cergy Paris Universit\'e, 95302 Cergy-Pontoise Cedex, France}

\date{\today}

\begin{abstract}
We consider a system of classical hard rods or billiard balls in one dimension, initially prepared in a Bragg-pulse state at a given temperature and subjected to external periodic fields. We show that at late times the system always thermalises in the thermodynamic limit via a 3-stages process characterised by: an early phase where the dynamics is well described by Euler hydrodynamics, a subsequent where a (weak) turbulent phase is observed and where hydrodynamic gradient expansion can be broken, and a final one where the gas thermalises according to a viscous hydrodynamics. As the hard rod gas shares the same large-scale hydrodynamics as other quantum and classical integrable systems, we expect these features to universally characterise all many-body integrable systems in generic external potentials.  
\end{abstract}

\maketitle

\prlsection{Introduction}
Interacting many-particle systems are notorious for bringing up new emergent physical behaviours at large scales that would be impossible to observe at the level of their elementary constituents. Understanding and classifying the emergent laws of non-equilibrium many-body dynamics is, therefore, one of the main focuses of present-day physics. In the past years, a lot of effort has been put on understanding the equilibration and thermalisation mechanism of isolated many-body interacting systems, at theoretical \cite{rigol_2006, rigol_2007,rigol2008thermalization,RevModPhys.91.021001,eisert_quantum_2014} and experimental level \cite{Kinoshita_quantum_2006, Bloch_exp_2008, Tang_exp_2018, schemmer_generalized_2018,PhysRevLett.126.090602,Wei2022,Malvania2021,Le2023}. Indeed, while the postulates of statistical physics indicate that eventually, the system thermalises, or at least it explores equally all the phase space at its disposal, much less is known about the approach to equilibrium, especially with regard to quantum systems \cite{deutsch1991,srednicki1994chaos,rigol_2007,essler_quench_2016}. Violation of canonical thermalisation on the other hand, can be observed in the presence of extra symmetries, for example, local symmetries, or extensive ones, as it is the case for localised or integrable systems \cite{Altman2018,srednicki1994chaos,vidmar2016generalized, essler_quench_2016}. 
There, despite conserved quantities are typically broken at large times in real settings, the dynamics of quasi-integrable systems is of extreme interest, as numerous experimental settings are much better described by integrable systems than fully non-integrable ones  \cite{Kinoshita_quantum_2006, Bloch_exp_2008, Tang_exp_2018, schemmer_generalized_2018, PhysRevLett.126.090602,Wei2022,Malvania2021}. Global integrability is typically broken by other Hamiltonian terms, such as the harmonic trap in a cold atomic setting, or an inhomogeneous magnetic field. As a full quantum problem is hard to numerically simulate for a long time, it is advisable to introduce toy models that are expected to display the same physics as the real quantum ones.

One of them is the \textit{hard rod gas}, namely billiard balls with a fixed diameter $d$ in one dimension that scatters elastically at each collision. The model is clearly integrable \cite{spohn_large_1991, Boldrighini_1, Percus_HR_1969} as all the initial momenta of the rods are preserved by dynamics, and it has been shown \cite{spohn_large_1991,doyon_drude_2017} to be described at large scale by the same hydrodynamics characterizing generic integrable systems \cite{Doyon_HR_2017, doyon_soliton_2017}, i.e. generalised hydrodynamics (GHD) \cite{castro-alvaredo_emergent_2016,bertini_transport_2016} (see also various applications to cold atomic as well as spin and fermionic systems \cite{bulchandani_bethe-boltzmann_2018,de_nardis_hydrodynamic_2018,bastianello_generalized_2017,doyon_soliton_2017,ilievski_ballistic_2017,bulchandani_classical_2017,fava_hydrodynamic_2021,bertini_low-temperature_2018,pozsgay_current_2020,PhysRevLett.123.130602,PhysRevB.102.161110,PhysRevB.103.165121, PhysRevLett.122.240606}). Moreover, the dynamics with integrability breaking terms in the hard rod gas has been the subject of numerous studies recently \cite{VasseurBackscattering,Olla} as indeed it is expected to display many similarities with the one observable in quantum gases. 

In this letter, we study a model of hard rods in the presence of {external inhomogeneous potentials}. We consider both a trapping potential $V(x)$ and a spacial dependent mass $m(x)$. While the latter may sound artificial, it actually mimics the dynamics of quasiparticles in spin chains under inhomogeneous magnetic fields (as the magnetic field gives an effective mass to the magnonic degrees of freedom, see for example \cite{PhysRevB.106.134314,Senaratne2022}). We here show that the system thermalises as long as external fields are finite and inhomogeneous by means of an effective \textit{viscous hydrodynamics}.  We observe that the thermalisation dynamics can be split into three main phases in time: I) an \textit{Euler phase} where the dynamics almost follow the one of the Euler GHD equation, II) a \textit{turbulent phase}, in close agreement with 2-dimensional wave turbulence \cite{Zakharov1992} phenomenology, where the state of the system deviates from the viscous hydrodynamics prediction and where the Fourier transform of the distribution of momenta shows a power law decay and III) a final \textit{diffusively thermalising phase} where the thermal state is approached exponentially with a timescale set by the diffusion constant of the final thermal state, fixed by the viscous GHD. While in phase I and III,  viscous GHD equation correctly capture \textit{quantitatively} the dynamics of the gas, it can fail to give correct predictions in phase II, due to a turbulence-induced \textit{gradient catastrophe}. However, if the initial momentum distribution is smooth, as in the case of initial high temperature, the effect of such a turbulent phase is reduced and viscous GHD captures the whole thermalisation dynamics.    

\prlsection{The hard rod gas}
We consider a system of $N$ billiard balls of diameter $d$ on a circle of length $L$ with Hamiltonian given by 
\begin{equation}\label{eq:Hamiltonian}
    H = \sum_{i=1}^N \Big[ \frac{ \theta_i^2}{2 m(x_i) }   + V(x_i) \Big] + \sum_{i<j} U(x_i-x_j),
\end{equation}
where $\theta_i^2$ is the momentum of each particle with its centre positioned in $x_i$, and where the interaction potential is the one describing hard spheres (Tonks gas), $U(x) = \infty$ for $|x|\leq d$ and $U(x) = 0$ for $|x|> d$. In this letter we consider time-independent periodically varying external potentials, with a wavelength $\ell$, namely, we consider 
\begin{equation}
    m(x) =  m  \Big( 1 + \frac{m_0}{m} {\cos (2 \pi x/\ell)} \Big)^{-1} ; \,
    V(x) =   V_0 \cos (2 \pi x/\ell),
\end{equation}
and we set $m=1$ generically. 
\begin{figure}[t]
\includegraphics[width=.45\textwidth]{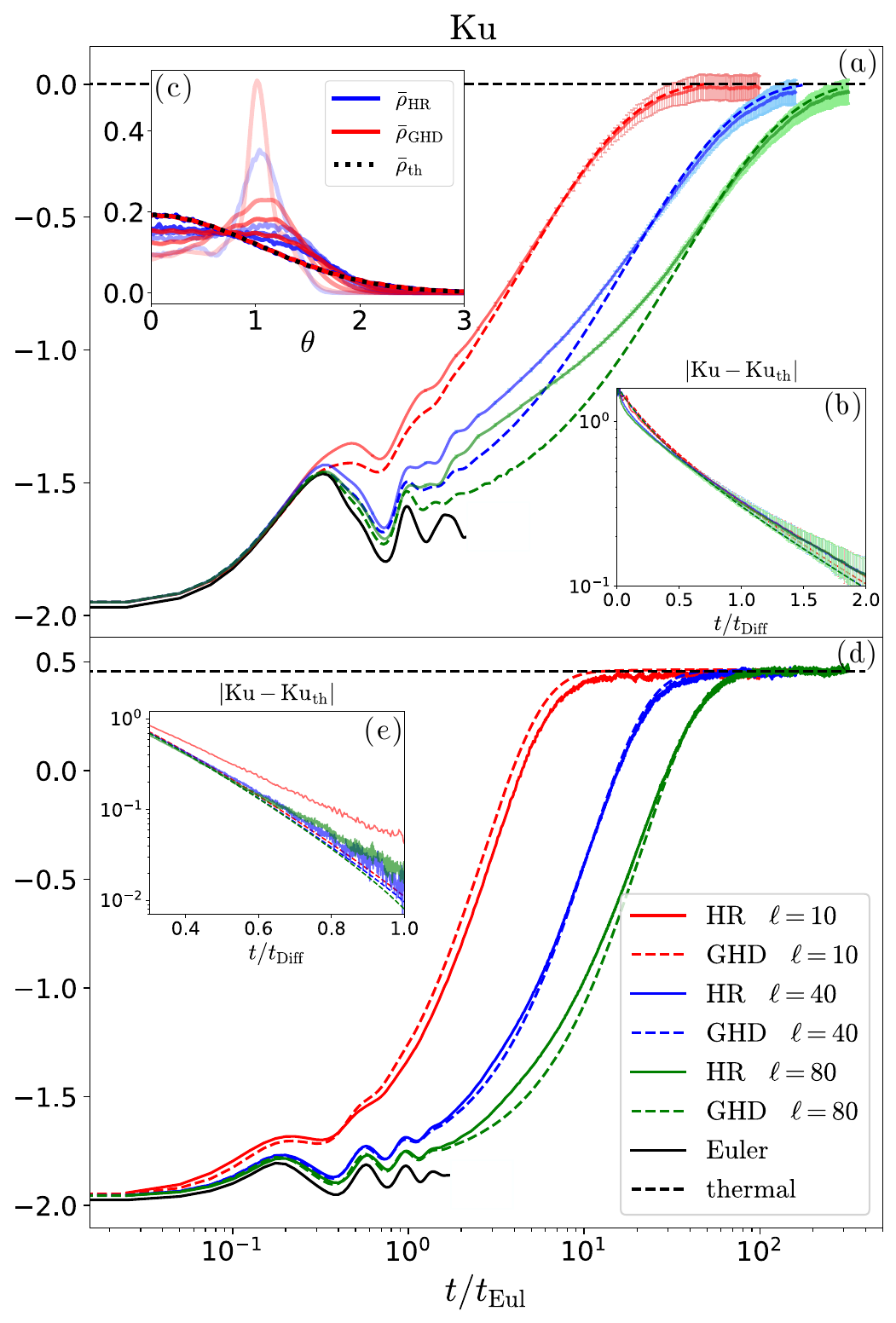}
\caption{Comparison of kurtosis of the spacially integrated $\rho(\theta, x ,t)$ of the exact hard rod gas dynamics (HR) with $d=1$,  and its diffusive GHD prediction, eq. \eqref{eq:hydroeq} (the inviscid, Euler hydrodynamics at short times is also reported in black),  with (a,b,c) $V_0 = 0.5,m_0=0$ and (d,e) $V_0 = 0,m_0=0.5$, initialised in a Bragg pulse state with low temperature $T_{0}=0.01$. Plots a,d show time evolution as a function of $t/t_{\rm Eul}$ and the insets b,e the approach to thermalisation as a function of $t/t_{\rm Diff}$. Plot c shows the spatially integrated $\rho(\theta, x ,t)$ for both GHD prediction and exact hard rod gas dynamics for $t/t_{\rm Eul}=0.5,5,10,20,160$ (increasing from light to dark) and the expected spatially integrated thermal distribution.    }
\label{fig:fig1}
\end{figure}
The Hamiltonian \eqref{eq:Hamiltonian} with $V_0 = m_0 = 0$ is integrable and is well studied in the literature \cite{PhysRev.171.224,Spohn1982,PhysRev.50.955,Robledo1986,PhysRevE.90.012147}. As scatterings are all elastic, the initial values of the momenta $\theta_i$ of the rods are conserved, giving $N$ integrals of motions. 
A generic stationary state of the integrable model, with zero external fields, is given by a so-called Generalised Gibbs ensemble (GGE) \cite{vidmar_generalized_2016}, namely a thermodynamic state where entropy is maximized given a distribution of momenta $\rho(\theta)$. One can simply use the mapping from free particles to hard-core ones to initialise the system in such a state, in particular, the spatial distribution of the particles is taken to be a Poisson point process \cite{Boldrighini_1, Boldrighini_2}. Such an initial state, when averaged over different realisations, has all the correct statistical properties of a GGE, i.e. the average number of particles in an interval $\Lambda$ is $\langle d N(\theta,\Lambda) \rangle =  \rho(\theta ) \Lambda$ and their fluctuations are given by the correct susceptibility matrix (for its explicit definition see, for example, \cite{doyon_drude_2017}) $C({\theta,\theta'}) \Lambda = \langle d N(\theta,\Lambda) d N(\theta',\Lambda) \rangle^c$. 
In this work, simulations are performed always on a ring of length $L=10^3$ , with $N=L/2$ particles and by averaging over $3\times 10^3$ initial configurations.
%

\prlsection{Viscous GHD}
The hard rod gas, despite being a purely classical model, shares exactly the same large-scale hydrodynamics with other quantum and classical integrable models, i.e. the GHD, where the distribution of the rods momenta $\rho(\theta, x ,t )$ at position $x$ and time $t$ is taken as the effective hydrodynamic fluid density. In the purely integrable case, there is no flow in the $\theta$ space, as all the momenta are conserved. However, in the presence of external forces, the flow in $\theta$ space gets activated. 
Indeed, switching on external potentials generically leads to integrability breaking. The momenta are not conserved any more and the resulting equations of motions for the rods, whenever the rod $i$ is not in contact with any of the other rods, result in $ \dot{x}_i(t) = 
 \frac{\theta_i(t)}{m(x_i)} $ and $\dot{\theta}_i(t) = a_i(t)$, 
with acceleration $a_i(t) =  \mathfrak{f}_1(x_i) + {\theta^2_i}  \mathfrak{f}_2(x_i)/2 $, where we introduced the forces $\mathfrak{f}_1(x) = - \partial_x V$ and $\mathfrak{f}_2(x) = - \partial_x \big(m(x)^{-1}\big)$. Including exact viscous dissipative terms, the relevant hydrodynamic equation takes the form of a two-dimensional fluid in the effective space $(\theta,x)$, with associated gradient $\nabla = \begin{pmatrix} \partial_x, & \partial_\theta \end{pmatrix}$, flow vector $\vec{J}^{\rm eff}_{[\rho]} = \begin{pmatrix} v^{\rm eff}_{[\rho]}, &  a^{\rm eff}_{[\rho]} \end{pmatrix} $ and diffusion matrix $ \boldsymbol{\mathfrak{D}}_{[\rho]} $ (whose definition is given in eq. \eqref{eq:diffusionmatrix}), 
reading as 
\begin{equation}\label{eq:hydroeq}
    \partial_t \rho + \nabla \cdot (\vec{J}^{\rm eff}_{[\rho]} \rho) = \frac{1}{2}  \nabla  \cdot (\boldsymbol{\mathfrak{D}}_{[\rho]} \nabla {\rho}).
\end{equation}
This equation is conjectured to describe generic integrable systems under external fields, and it was first derived in full generality in \cite{durnin_inhomogeneous_2021}.  In \cite{SM} is given an intuitive derivation in the case of the hard rod gas only based on the kinetic picture, in the same spirit as \cite{PhysRevB.98.220303,gopalakrishnan_kinetic_2018}. 

\begin{figure*}[t]
    \centering
    \includegraphics[width=0.9\textwidth]{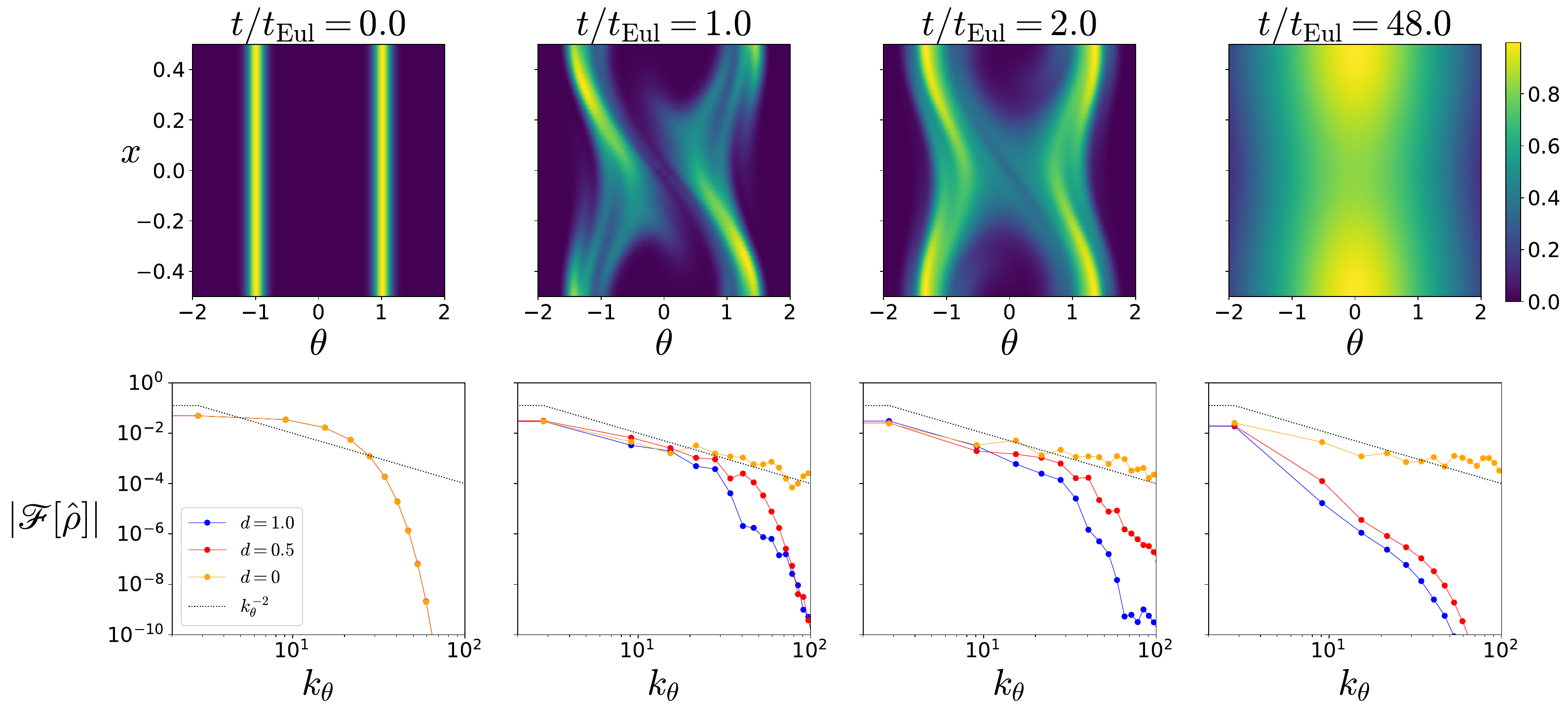}
    \caption{ GHD evolution, eq. \eqref{eq:hydroeq}, with $m_0=0$ and $V_0 = 0.5$ and $\ell=80$, initialised in a Bragg pulse state with $T_{0}=0.01$: (top) density plot of the density $\rho(\theta,x,t)$ as function of momentum $\theta$ and position $x$ at different times, for rods with length $d=1$. (bottom) Log-Log plot of the absolute value of the Fourier transform in $\theta$ of the spatially integrated density as a function of Fourier vector $k_\theta$ and three different values of the rods' length $d$. At time $t=0$ the decay is exponential while within the turbulent phase $t_{\rm Eul} \lesssim  t \lesssim t_{\rm Diff}$, we observe the decay as $k_\theta^{-\alpha}$ with $\alpha \sim 2$ in a range of $k_\theta \lesssim 1/\Lambda_{\rm sink} $ with $\Lambda_{\rm sink}  \sim \bar{\rho} d^2$. The plot for the mass field $m_0=0.5$ and $V_0 = 0$, as well as with the exact hard rods simulations, is in \cite{SM}.  }
\label{fig:fig2}
\end{figure*}

In order to define the flow vector and the diffusive matrix, it is useful to move from the elementary rods to the quasiparticle tracers. The tracers are defined as labels associated with the rods. During a scattering, the two rods exchange the tracer label. Such tracers, therefore, move throughout the system, experiencing jumps in space and momenta at each scattering with other particles.
Therefore, at the mesoscopic scale, the phase space dynamics depends on the local density $\rho(\theta,x)$ and is described by the flow vector 
\begin{equation}
    \vec{J}^{\rm eff}_{[\rho]}= \begin{pmatrix} 
    v^{\rm eff}_{[\rho]} =  (\theta  -  \overline{\rho\theta }d )/\big(m(x)(1-  \bar{\rho} d)\big)
    \\
    a^{\rm eff}_{[\rho]} = \mathfrak{f}_1+\mathfrak{f}_2(\theta^2/2  -  \overline{\rho\theta^2 }d/2 )/(1-  \bar{\rho} d)
    \end{pmatrix} .
\label{eq: effective vel and acc}
\end{equation}
where over-lined quantities denote integral over momenta $\bar{f}= \int {\rm d}\theta f(\theta)$.
Since the hydrodynamic flow \eqref{eq: effective vel and acc} depends non-trivially on the local particle density $\rho(\theta,x)$, it is subjected to a diffusive broadening due to its local finite thermodynamic fluctuations defined by the susceptibility matrix $C({\theta,\theta'})$. Notice that in non-integrable systems such quadratic fluctuations of convective currents give generically only a lower bound to the diffusion constants \cite{PhysRevE.89.012142}, while in integrable ones they fully saturate them \cite{Doyon2022}. Here we show that the latter remains true also in quasi-integrable systems. 

The diffusion coefficients $\boldsymbol{\mathfrak{D}}_{[\rho]}$ are then defined as generators of reversible Markov jump processes induced by such density fluctuations \cite{spohn_large_1991, Doyon_HR_2017} and are equal to 
\begin{gather}\label{eq:diffusionmatrix}
   \boldsymbol{\mathfrak{D}}_{[\rho]} =  \delta_{\theta,\theta'}\int_{\mathbb{R}}{\rm d}\theta'' \rho(\theta',x')\vec{g}(\theta,\theta'') - \rho(\theta',x)\vec{g}(\theta,\theta') ,
\end{gather}
where $\delta_{\theta,\theta'}$ is the Dirac delta in the momenta space and the kernel matrix $\vec{g}$ is defined by
\begin{equation}
  \frac{ \vec{g}}{d^2}=   
   \begin{pmatrix}
       |v^{\rm eff }_{[\rho]}(\theta)-v^{\rm eff }_{[\rho]}(\theta')| 
       & 
       \xi_{\theta,\theta'}\big(a^{\rm eff }_{[\rho]}(\theta)-a^{\rm eff }_{[\rho]}(\theta')\big)
       \\
       \xi_{\theta,\theta'}\big(a^{\rm eff }_{[\rho]}(\theta)-a^{\rm eff }_{[\rho]}(\theta')\big)
       &
       \frac{\big(a^{\rm eff }_{[\rho]}(\theta)-a^{\rm eff }_{[\rho]}(\theta')\big)^2}{|v^{\rm eff }_{[\rho]}(\theta)-v^{\rm eff }_{[\rho]}(\theta')|}
       \\
   \end{pmatrix}
   ,
\label{eq: g matrix}
\end{equation}
with $\xi_{\theta,\theta'}\equiv {\rm sgn}\big(v^{\rm eff}_{[\rho]}(\theta)-v^{\rm eff}_{[\rho]}(\theta')\big)$ (for a full derivation from kinetic theory see \cite{SM}).
While the left-hand side of \eqref{eq:hydroeq}, which corresponds to the Euler GHD equation, has numerous integrals of motion \cite{10.21468/SciPostPhys.6.6.070,cao_incomplete_2017} and does not thermalise, one can show (see \cite{SM}) that the viscous terms on the right-hand side break most of the conservation laws, and the fixed point of eq. \eqref{eq:hydroeq} is a thermal LDA (local density approximation) state, (as discussed also in \cite{PhysRevLett.125.240604,durnin_inhomogeneous_2021}) determined only by the inverse temperature $\beta$ and chemical potential $\mu$, fixed by the only conserved quantities: the integrated density $ N_0 =  \int {\rm d}x \int {\rm d}\theta  \rho(\theta,x,0) $ and energy $E_0 = \int {\rm d}x \int {\rm d}\theta  \rho(\theta,x,0)  ( \theta^2/(2m(x)) + V(x)) $.  In terms of these parameters, the thermal distribution reads  
\begin{equation}{\label{eq:thermaldistribution}}
\rho_{\rm th}(\theta; x ) \sim  e^{- \beta( m(x) \theta^2/2 + V(x) - \mu)} .
  \end{equation}
 We here shall consider the integrated distribution in space 
$\hat{\rho}(\theta,t) = \int {\rm d}x   \rho(\theta, x, t)$
and show that \textit{at late times thermalisation is always achieved} $ \hat{\rho}(\theta,t) \to  \hat{\rho}_{\rm th}(\theta)$ under the effect of the external forces and internal viscosities, by means of the mechanism we describe below.

\prlsection{The three-stage thermalisation}
We now focus on a concrete example of dynamics from an initial state. We mimic the state created by a strong Bragg pulse in cold atomic gas, as done for example in \cite{Kinoshita_quantum_2006,Le2023}. In the context of the hard rods, we prepare a spatially homogenous gas with a distribution of momenta given by two Gaussian peaks with a given initial temperature  $T_{0}$
\begin{equation}\label{eq:initialstate}
  \rho(\theta, x , t=0) \sim    e^{- \frac{(\theta-\theta_{\rm Bragg})^2}{2T_0}  }     + e^{- \frac{(\theta+\theta_{\rm Bragg})^2}{2T_0}  }         ,
\end{equation}
 with normalisation fixed by density $\bar{\rho}=1/2$ and with $\theta_{\rm Bragg} = 1$ here for convenience. We then let the system evolve under an external field, either with $V_0$ or $m_0$ finite. We solve the GHD equation \eqref{eq:hydroeq} by means of a combined 4th order Runge-Kutta method as defined in \cite{2212.12349} and an implicit mid-point method and by rescaling space $x$ and time $t$ by $\ell$, which allows solving it in $x \in [0,1]$, by adding a factor $1/\ell$ in front of the diffusive terms. We compare the GHD prediction with exact numerical simulations of hard rods, see \cite{SM} for details on the algorithm. We focus on the integrated density $\hat{\rho}(\theta)$ (which can be obtained by a trap-release protocol in a cold atomic setting \cite{Malvania2021,bouchule2023}) and in particular on its kurtosis ${\rm Ku}(t) = \bar{\theta^4}/(\bar{\theta}^2)^2 -3$ with the overline average taken with $\hat{\rho}(\theta, t )$ as a measure, see Fig. \ref{fig:fig1}.  When $m_0 =0$, thermalisation is manifested by ${\rm Ku}(t) \to 0$ at large times, signalling that the integrated distribution converges to its thermal value \eqref{eq:thermaldistribution}. However, when $m_0 \neq 0$ the integrated thermal distribution is not a Gaussian, as can be seen from expression \eqref{eq:thermaldistribution} and a different value of non-zero kurtosis is therefore expected at large times. For values of $\ell$ relatively large to ensure the validity of the fluid cell approximation, at times $t \lesssim t_{\rm Eul}\equiv \ell m/\theta_{\rm Bragg}$ we always observe agreement between numerical hard rods simulations and viscous GHD, where the latter can also be simplified to its Euler formulation with no dramatic change of behaviour, see Fig. \ref{fig:fig1} where the Euler GHD prediction is reported (up to the time after which the numerical simulation becomes unstable). We denote this phase as the Euler phase, where the dynamics is purely given by an adiabatic Euler flow. As soon as time becomes comparable with $\sim  t_{\rm Eul}$ something dramatic happens: when the gas initial temperature is small $T_0/(\theta_{\rm bragg}^2) \ll 1$, the exact numerical simulations and the hydrodynamic predictions do not agree any more, even for large $\ell$ (actually larger is $\ell$, stronger are the deviations from hydrodynamics). We denote this phase as the \textit{turbulent phase}. This phase is characterised by a proliferation of discontinuities in the derivative of the momentum distribution $\rho(\theta, x, t)$, causing the hydrodynamic gradient expansion to break down (gradient catastrophe), whenever the gas is interacting (a similar mechanism for free fermion gases was discussed in \cite{PhysRevA.98.043610}). In order to quantify such a discontinuous behaviour, we consider the absolute value of the Fourier transform in $\theta$ of the integrated momenta distribution $|\mathfrak{F}[\hat{\rho}](k_\theta)| =  |\int {\rm d}\theta e^{i k_\theta \theta}\hat{\rho}(\theta, t)|/2\pi$, see Fig. \ref{fig:fig2}. The latter decays exponentially in $k_\theta$ at short times (being the Fourier transform of two Gaussian peaks) to then develop a power law decay as $k_\theta^{-2}$, compatible with the presence of cusps. For momenta larger than a dissipation scale $k_\theta \gtrsim 1/\Lambda_{\rm sink} $, (the latter fixed by the viscosity  $\Lambda_{\rm sink}  \sim \boldsymbol{{\mathfrak{D}}}/\theta_{\rm Bragg} \sim d^2  \bar{\rho}$ and by the initial temperature $T_0$, the power law decay transients to an exponential one, as expected from typical turbulent spectra. As the system under consideration here is not driven, eventually viscous terms take over as time progresses and redistribute smoothly momenta in the $(\theta,x)$ space. At times $t \sim t_{\rm Diff}$, with (see \cite{SM} for a derivation)
\begin{align}
& t_{\rm Diff}\equiv \ell^2 \ m\theta_{\rm Bragg}/(d^2\bar\rho V_0), \nonumber  \\& 
t_{\rm Diff}\equiv  \ell^2\ m^2(m-m_0)^2/(m_0^3 d^2\bar\rho\theta_{\rm Bragg}), 
\end{align}  
   respectively in the presence of space-dependent trapping potential or space-dependent mass  (see also \cite{SM} for rescaled plots in these units), there is no trace of turbulent behaviour any more and the system thermalizes by redistributing a remaining low-momentum density ripple (in $x$ direction) on a timescale set by the minimal diffusion constant in the system. In this phase, denoted as \textit{diffusively thermalising phase}, hydrodynamics is restored and gives quantitatively correct prediction for the decay time and the momenta distributions. To clarify such mechanism we show also the free particle case $d=0$ in Fig. \ref{fig:fig2} (see also \cite{SM} for additional data on this case):  the same proliferation of discontinuities in the single particle density $\hat{\rho}(\theta)$ occurs, but being Euler hydrodynamics there an exact description of the system, with no higher derivative terms, no breaking of hydrodynamics expansion occurs. In the interacting case $d>0$ instead, Euler and diffusive GHD are approximate theories and higher order derivatives in $x$ and $\theta$ are generically present, and they are expected to become relevant at times $t/t_{\rm Eul} \sim 1$. {Therefore, the inset of the turbulent phase does not signal the breakdown of Euler GHD but rather the \textit{breakdown of Hydrodynamic gradient expansion}, so that viscous GHD fails to provide a complete description of the state. }

\begin{figure}[t]
\includegraphics[width=.43\textwidth]{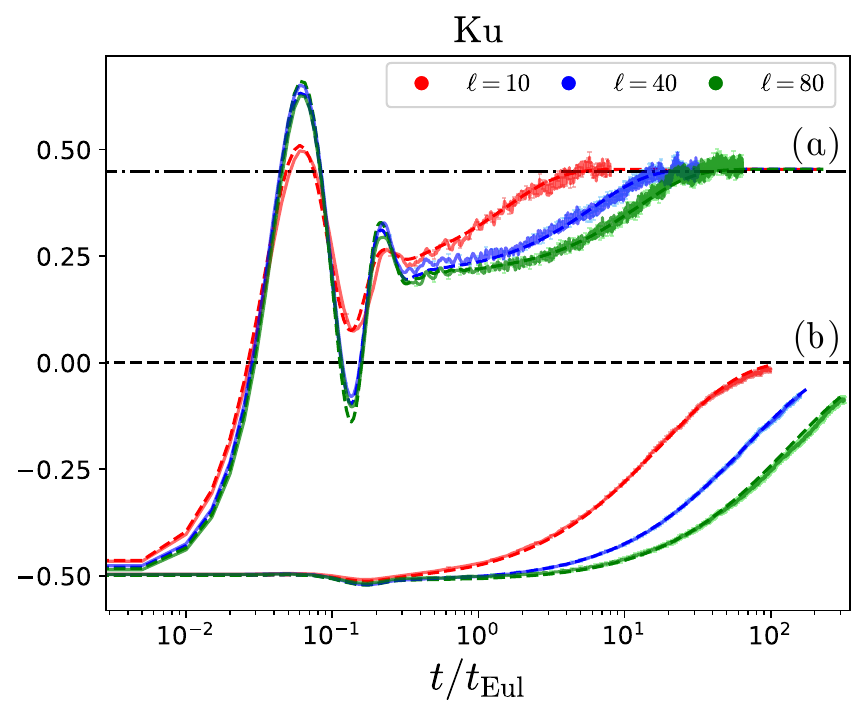}
\caption{Comparison of kurtosis of the spacially integrated $\rho(\theta, x ,t)$ of the exact hard rod gas dynamics (HR) with $d=1$,  and diffusive GHD prediction for a larger initial temperature $T_0=1.0$. Line (a) is the kurtosis value of the thermal spatially integrated $\bar{\rho}_{\rm th}$ with $V_0 = 0,m_0=0.5$ and line (b) with $V_0 = 0.5,m_0=0$. Coloured dashed lines correspond to the GHD predictions and solid lines for the exact hard rod dynamics.}
\label{fig:fig3}
\end{figure}

Finally, we emphasise that the relevance of such a turbulent regime depends on the initial state. Initial states of the type of eq. \eqref{eq:initialstate} with larger $T_0/\theta_{\rm Bragg}^2$ ratio, have larger $\Lambda_{\rm sink}$ and thus a less pronounced turbulent phase and diffusive GHD well reproduces the dynamics at all times, see Fig. \ref{fig:fig3}.  Moreover, we also stress the surprising consequence of our results: the GHD description is actually more accurate when $\ell$ is small, i.e. for strongly broken integrability, as compared with large $\ell$, i.e. for weakly broken integrability. Clearly, if $\ell$ is taken to be very large, the timescale for turbulence  $t \sim t_{\rm Eul}\propto\ell$ and for thermalisation  $t \sim t_{\rm Diff}\propto \ell^2$ does increase, but the turbulence phase will be even stronger as $\ell$ is increased, as diffusion needs longer times to fully redistribute momenta throughout the system. 

\prlsection{Conclusion}
In this letter, we have studied how under the influence of external fields, a system of interacting hard rods thermalises, a question that has received contradictory answers in the past few years \cite{cao_incomplete_2017,Vir2023}, and which is extremely relevant in order to understand the thermalisation mechanisms for quasi-integrable models.

Our findings are summarised as the following: in the case of smooth (in momentum space) initial states, the evolution of the distribution of momenta remain also smooth both in space and momentum direction, and viscous GHD provides an exact description of the thermalisation dynamics. When instead the distribution develops cusps, a turbulent phase kicks in and the hydrodynamic gradient expansion fails: viscous GHD is not enough to predict the whole thermalisation dynamics, yet it provides a good description of the system at early and, surprisingly enough, late times.
Therefore, quite counter-intuitively, viscous GHD predicts more accurately the behavior of a gas with strong external forces rather than one with weak ones.

Our work opens new unexplored and exciting directions, in particular on the nature of turbulence and hydrodynamic breaking and restoration in quasi-integrable systems.
Indeed, as their effective hydrodynamics is two-dimensional,  weaker forms of turbulence, as for example the wave turbulent phase, recently observed in a two-dimensional weakly interacting Bose gas \cite{Navon2016,Navon2019,Galka_turbolence_2022} and in other quasi-integrable systems in \cite{PhysRevLett.129.114101}, are therefore possible. This paper provides a first example and open the way for new exiting developments for example within the Lieb-Liniger gas or the Gross-Pitaevskii equation \cite{Koch_2022,bouchule2023,10.21468/SciPostPhys.9.1.002}, which would open the way to a possible experimental observation of such a turbulent behaviour. 

\prlsection{Acknowledgements}
We are grateful to B. Doyon for useful discussions and for explaining how to numerically prepare hard rods GGE states. We thank A. Bastianello, A. De Luca, S. Pappalardi for enlightening discussions and related collaborations.  This work has been partially funded by the ERC Starting Grant 101042293 (HEPIQ).

\bibliography{biblio}

\onecolumngrid
\newpage

\appendix
\setcounter{equation}{0}
\setcounter{figure}{0}
\renewcommand{\thetable}{S\arabic{table}}
\renewcommand{\theequation}{S\thesection.\arabic{equation}}
\renewcommand{\thefigure}{S\arabic{figure}}
\setcounter{secnumdepth}{2}

\begin{center}
{\Large Supplementary Material \\ 
\titleinfo
}
\end{center}
\tableofcontents

\appendix

\begin{appendix}
\section{Additional numerical data}
\begin{figure}[H]
    \centering
    \includegraphics[width=1.0\textwidth]{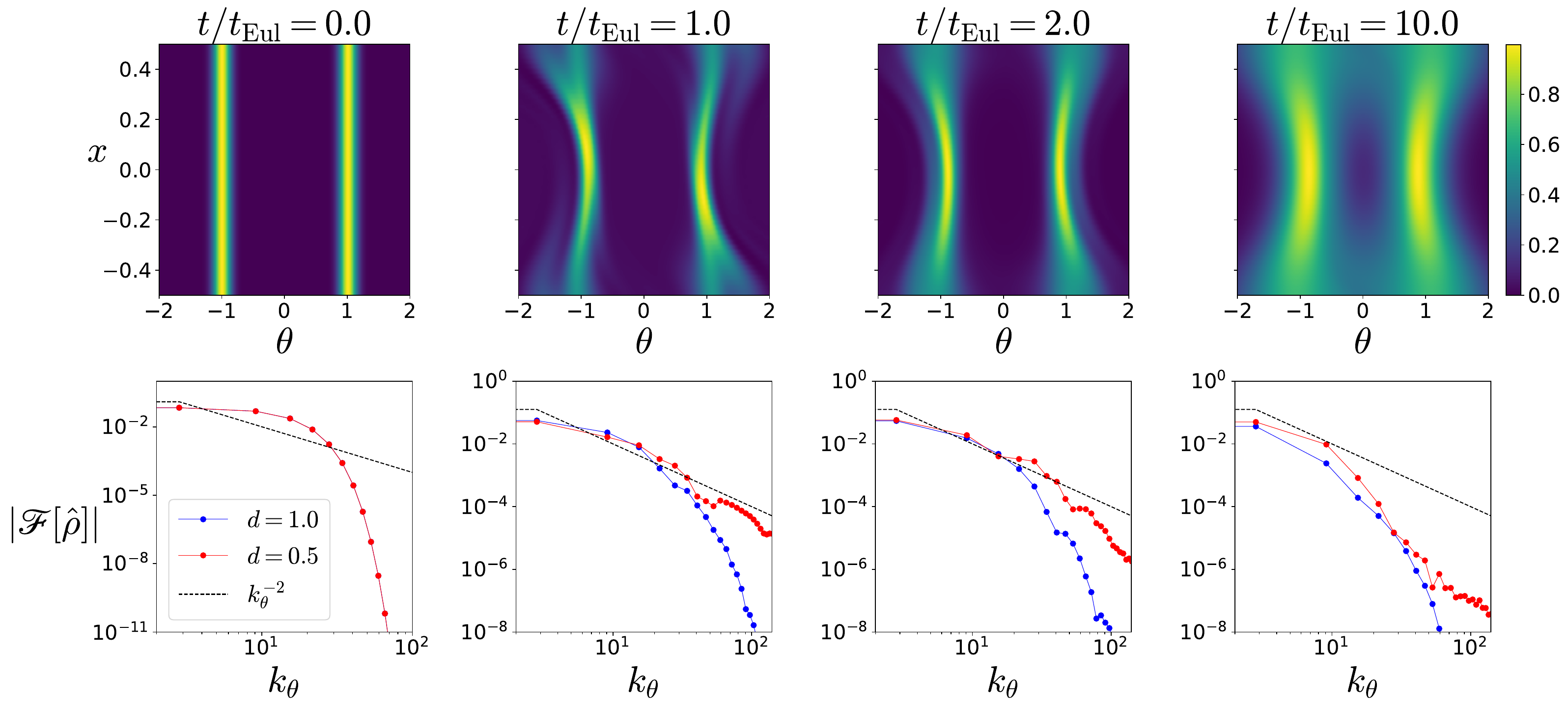}
    \caption{ GHD evolution with $m_0(x) = 0.5$ and $V_0 = 0$ and $\ell=80$, initialised in a Bragg pulse state with $T_0=0.01$: (top) density plot of the density $\rho(\theta,x,t)$ as function of momentum $\theta$ and position $x$ at different times, for rods with length $d=1$. (bottom) Log-Log plot of the absolute value of the Fourier transform in $\theta$ of the spacially integrated density as a function of Fourier vector $k_\theta$ and two different values of the rods' length $d$. The behaviour of the system, in this case, is qualitatively the same as the one shown in Fig. 2 of the main text.}
\label{fig:fig_FT_V2}
\end{figure}

\begin{figure}[H]
    \centering
    \includegraphics[width=1.0\textwidth]{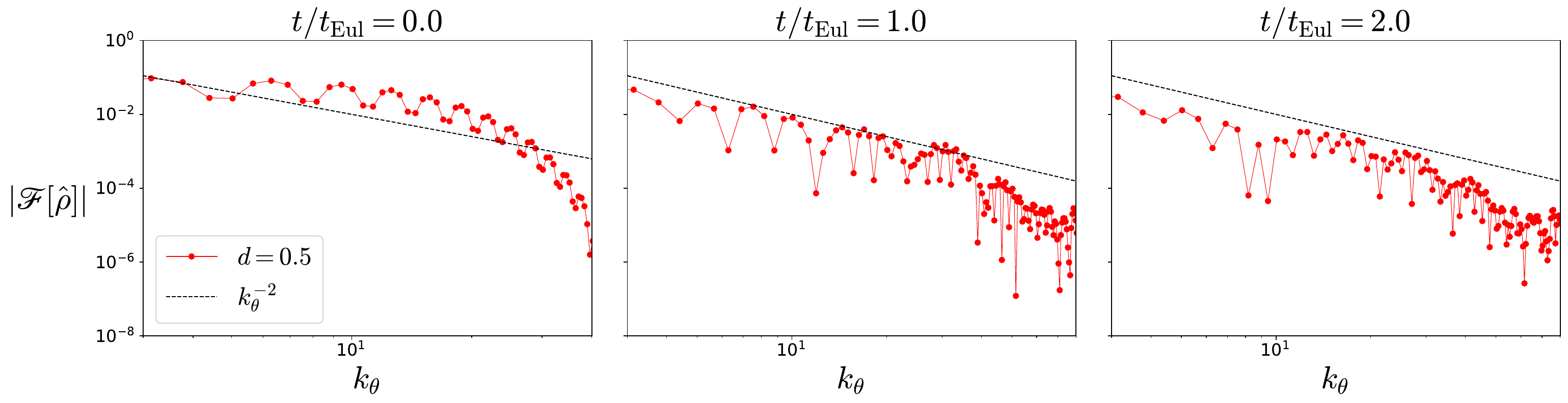}
    \caption{ Exact hard rods gas evolution with $m_0(x) = 0$ and $V_0 = 0.5$ and $\ell=80$, initialised in a Bragg pulse state with width $T_0=0.01$. Log-Log plot of the absolute value of the Fourier transforms in $\theta$ of the spacially integrated density estimated through exact numerical simulations of hard rods with rods' length $d=0.5$. 
    The data are shown as a function of Fourier vector $k_\theta$. The behaviour of the system, in this case, is qualitatively the same as the one shown in Fig. 2 of the main text, although we can resolve a smaller range of momenta compared to GHD simulations due to Monte Carlo noise. }
\label{fig:fig_FT_V1_HR}
\end{figure}

\begin{figure}[H]
    \centering
    \subfloat[][]
   {\includegraphics[width=0.45\textwidth]{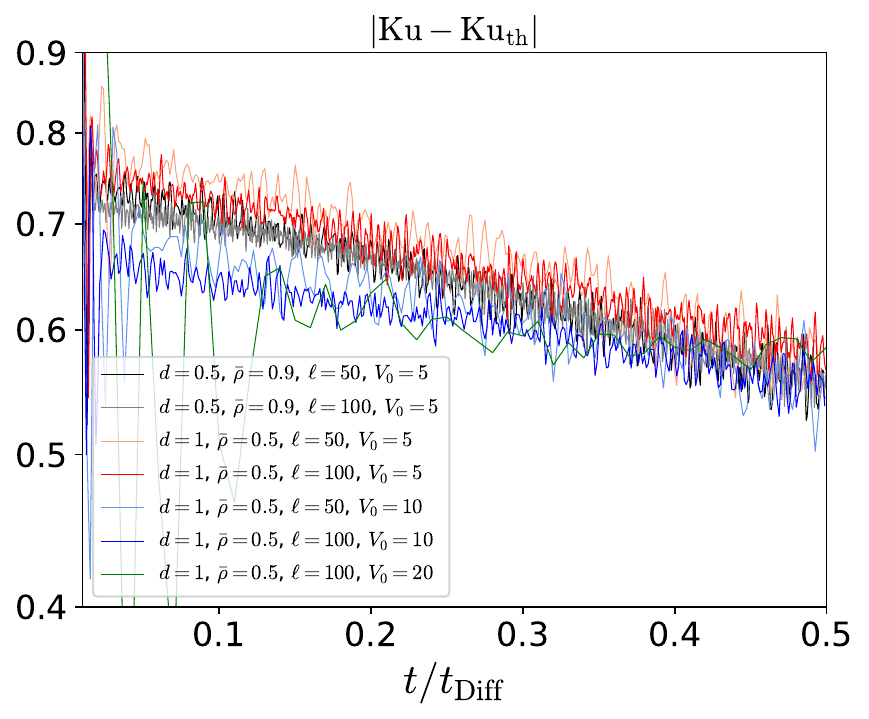}} \quad
    \subfloat[][]
   {\includegraphics[width=0.45\textwidth]{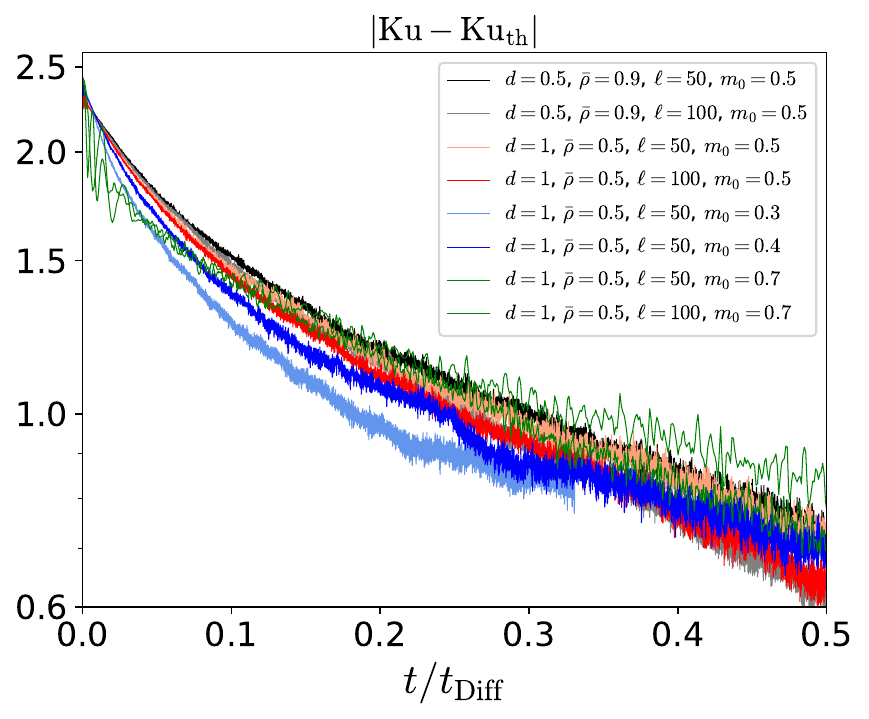}}
    \caption{Kurtosis of the spacially integrated $\rho(\theta, x ,t)$ of the exact hard rod gas dynamics with $d=1$, $\theta_{\rm Bragg }=1$, $T_0=0.01$, $\ell=L$ and (a) $m_0=0$, (b) $V_0=0$. The dynamics is shown for different values of $d$, $\bar\rho$, $L$ and  (a) $V_0$, (b) $m_0$. The Log plots show the time evolution as a function of the rescaled time (a) $t/t_{\rm Diff}\equiv t d^2 \bar\rho V_0 /(\theta_{\rm Bragg }\ell^2)$  and (b) $t/t_{\rm Diff}\equiv t d^2 \bar\rho \theta_{\rm Bragg}m_0^3 /(m^2(m-m_0)^2\ell^2)$ as expected from GHD equations. The collapse of the curves is in agreement with the definition of diffusive time scales presented in appendix \ref{sec:time scales}. }
\label{fig:Kurtosis_Scaling}
\end{figure}

In this appendix we present the following numerical results
\begin{itemize}
\item Fig.\ref{fig:fig_FT_V2}:  we show the absolute value of the Fourier transform in $\theta$ of the spacially integrated density given by the GHD evolution with $m_0=0.5$ and $V_0=0$. We observe that, in the presence of a space dependent mass, the  behaviour of this quantity is qualitatively the same as in the presence of a potential trap, shown in Fig.2 of the main text.
\item Fig.\ref{fig:fig_FT_V1_HR}: we show the absolute value of the Fourier transform in $\theta$ of the spacially integrated density given by the exact hard rods gas evolution with $V_0=0.5$ and $m_0=0$ . The results are qualitatively the same as in Fig.2 of the main text. Although GHD is expected not to be predictive in the turbolent phase, it mimics the correct qualitative behaviour for this quantity.
\item Fig.\ref{fig:Kurtosis_Scaling}: kurtosis of the spatially integrated density as function of the rescaled time $t/t_{\rm Diff}$ for many different values of $d$, $\bar\rho$, $\ell$, $V_0$ and $m_0$. The collapse of the curves is in agreement with the definition of diffusive time scales presented in appendix \ref{sec:time scales}.
\item Fig.\ref{fig:Kurtosis_d=0}: kurtosis of the spacially integrated density of the exact hard rod gas dynamics with $d=0$ and its  Euler scale GHD prediction. The plot show that the Euler scale GHD equation is exact at all times. In the limit $d=0$ of non-interactive particles, the system is integrable and the steady state is not thermal.
\item Fig.\ref{fig:fig_FT_V1_d0}: Euler scale GHD evolution for $d=0$ with $m_0(x) = 0$, $V_0 = 0.5$. n the limit $d=0$ of non-interactive particles, the Euler scale GHD equation is exact and it forecast a turbolent phase for $t \gtrsim t_{\rm Eul}$ with the same decay exponent as the one observed for $d>0$.
\end{itemize}

\newpage

\begin{figure}[H]
    \centering
    \includegraphics[width=0.5\textwidth]{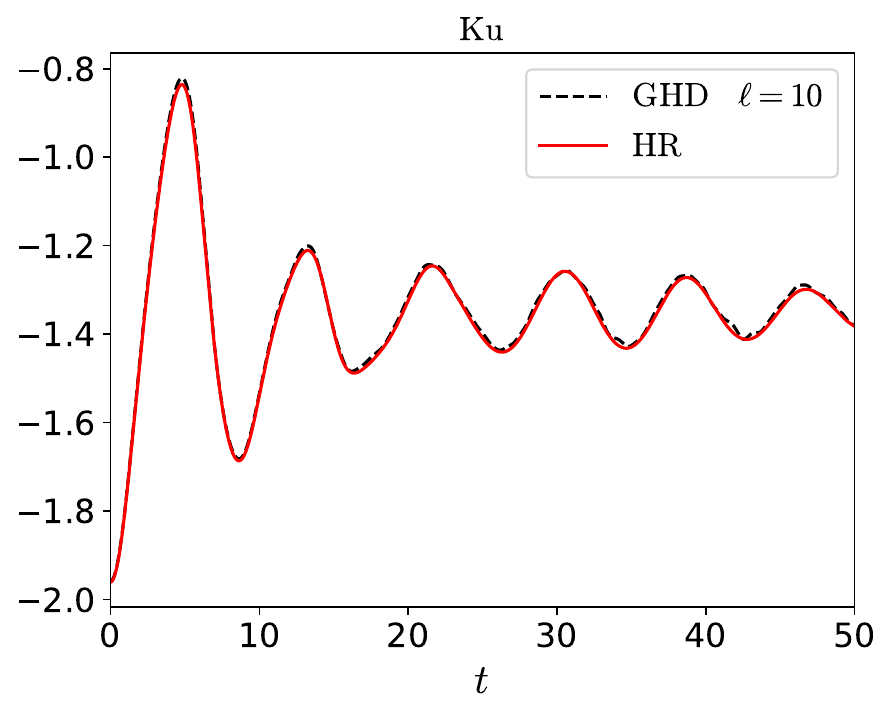}
    \caption{Kurtosis of the spacially integrated $\rho(\theta, x ,t)$ of the exact hard rod gas dynamics with $d=0$ and its GHD prediction with $V_0=0.5$, $m_0=0$ and $\ell=10$. The initial state is a Bragg pulse with $T_0=0.01$. In the limit $d=0$ of non-interactive particles, the rods are independent and Euler GHD describes exactly the phase space evolution of the system. Hence, the system is integrable and the steady state is not the thermal one. }
\label{fig:Kurtosis_d=0}
\end{figure}

\begin{figure}[H]
    \centering
    \includegraphics[width=1.0\textwidth]{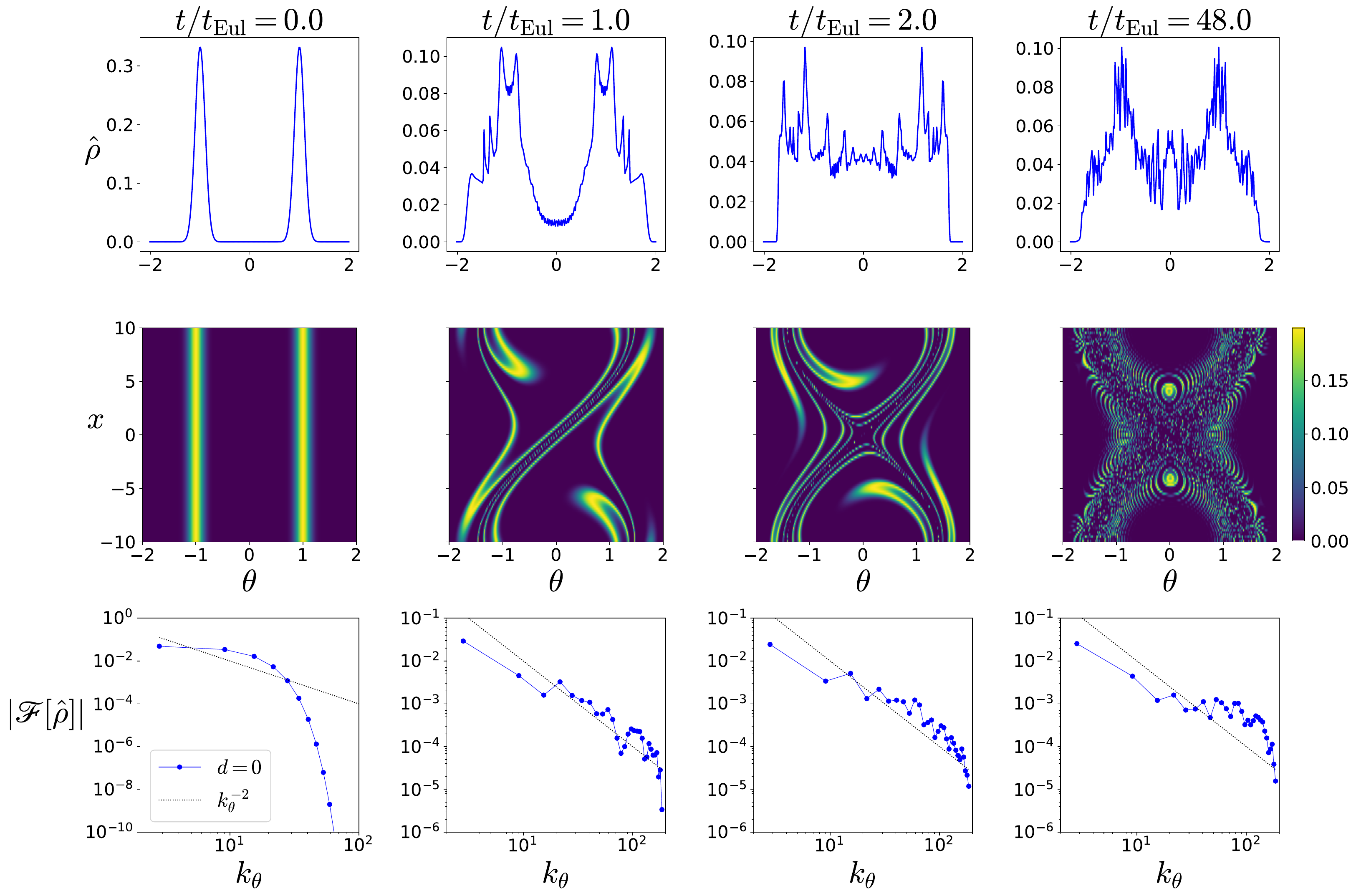}
    \caption{ Euler scale GHD evolution for $d=0$ with $m_0(x) = 0$, $V_0 = 0.5$ and $\ell=5$, initialised in a Bragg pulse state with width $T_0=0.01$: (top) Plot of the spacially integrated density as function of momentum $\theta$ at different times. (middle) Plot of the density as function of momentum $\theta$ and position $x$. (bottom) Log-Log plot of the absolute value of the Fourier transform in $\theta$ of the spacially integrated density as a function of Fourier vector $k_\theta$. At $t=0$ the decay is exponential while within the turbolent phase $t/\ell \gtrsim 1 $ we observe the (fitted) decay $k_{\theta}^{-\alpha} $ with $\alpha\sim 2$. We stress that, in the non-interactive particles limit $d=0$, the Euler scale GHD is exact at all times.  }
\label{fig:fig_FT_V1_d0}
\end{figure}

\section{The hard rods algorithm }
In this appendix, we describe the algorithm used to simulate the dynamics of a hard rod gas. 
We consider a system of $N$ particles $\{P_j, j=1,...,N\}$ having length $d$ and coordinates $\{(x_j,\theta_j), j=1,...,N\}$. They move freely in a one-dimensional space of length $L$, except for elastic collisions that conserve energy and momentum. The particles are also ordered from left to right $x_{j+1}\geq x_j+d$.
The single-particle Hamiltonian of the system is
\begin{equation}
    H(\theta,x)=\frac{\theta^2}{2m(x)}+V_1(x),
    \label{eq:1-particle Hamiltonian}
\end{equation}
where $V_1(x)$ is an external trapping potentials and $m(x)$ is a space dependent mass. The hard rods algorithm for the time evolution of the system is the following \cite{Alba_flea_gas_2020}:
\begin{enumerate}
    \setcounter{enumi}{-1}
    \item Initialize the system in a GGE state (see subsection \ref{sec:initilise}).
    \item \label{first step } Find the pair of particles $\{P_j,P_{j+1}\}$ that scatter first and their scattering time $t_{\rm coll}$.
    \item Let all the particles evolve without interaction until $t_{\rm coll}$.
    \item \label{laste step} Perform the scattering between particles $\{P_j,P_{j+1}\}$.
    \item Iterate \ref{first step }-\ref{laste step}.
\end{enumerate}
\subsection{Initialization of the system}\label{sec:initilise}
We discuss here how to initialize the rods in a GGE state, homogeneous in space and with an arbitrary momentum distribution $h(\theta)$.\\ 
In a hard rod GGE state, the distribution of particles in space is a Poisson point process \cite{Boldrighini_1, Boldrighini_2}. In particular, the probability of finding the edges of two neighbouring particles at distance $y$ is equal to
\begin{equation}
    \mathbb{P}(y)= \frac{\bar\rho}{1-d\bar\rho}{\rm e}^{-y\frac{\bar\rho}{1-d\bar\rho}},
    \label{eq: exponential distribution}
\end{equation}
with $\bar\rho=N/L$. In order to impose such statistics, we initialize the system with the following procedure: we place the particle with $j=1$ at distance $y_1+d$ from the left border, where $y_1$ is extracted from a distribution \ref{eq: exponential distribution}. We repeat for particles with $j>1$, placing them at position $x_j=x_{j-1}+y_j+d$ until we reach the right border. Then we extract the momentum of particles from the distribution $h(\theta)$.
It can be proven that the mean number of particles in the system is actually $N$. 

\subsection{Free particle dynamics and collisional time}
The equations of motion for the Hamiltonian defined in eq. \ref{eq:1-particle Hamiltonian} are
\begin{equation}
\begin{cases}
    \dot{x} = \frac{\theta}{m(x)}, \\
    \dot{\theta} = \mathfrak{f}_1(x) +  \frac{\theta^2}{2} \  \mathfrak{f}_2(x) ,
\end{cases}
\label{eq:eq motion}
\end{equation}
with $\mathfrak{f}_1\equiv-\partial_x V_1(x)$ and $\mathfrak{f}_2\equiv-\partial_x m^{-1}(x)$. Considering non-trivial potentials, the system \ref{eq:eq motion} of nonlinear differential equations is not solvable analytically. Thus, we use a second-order explicit Runge-Kutta method to compute the time evolution of a free particle.\\
We adopt a time discretization with time step ${\rm d}t$ and evolve the particle coordinates $\{x_{[n-1]},\theta_{[n-1]}\}\rightarrow\{x_{[n]}, \theta_{[n]}\}$ according to the equations
\begin{equation}
    \begin{split}
        x_{[n]} &= x_{[n-1]}+\frac{K_{v,n}}{m(K_{x,n})} {\rm d}t 
        ,
        \\
        \theta_{[n]} &= \theta_{[n-1]}+\big(\mathfrak{f}_1(K_{x,n})+\mathfrak{f}_2(K_{x,n})\frac{K_{v,n}^2}{2}\big){\rm d}t ,
        \\
        K_{x,n} &\equiv x_{[n-1]}+\frac{\theta_{[n-1]}}{m(x_{[n-1]})} \frac{{\rm d}t}{2}
        ,
        \\
        K_{v,n} &\equiv \theta_{[n-1]}+\big(\mathfrak{f}_1(x_{[n-1]})+\mathfrak{f}_2(x_{[n-1]})\frac{\theta_{[n-1]}^2}{2}\big)\frac{{\rm d}t}{2} 
        ,
    \end{split}
\label{eq: Runge-Kutta HR}
\end{equation}
where $\{x_{[n]}, \theta_{[n]}\}$ is the Runge-Kutta approximation of $\{x(t_n), \theta(t_n)\}$ with $t_n=n\,{\rm d}t$.
According to equation \ref{eq: Runge-Kutta HR}, the collisional time between two neighbouring particles $\{P_j,P_{j+1}\}$ is
\begin{equation}
    t^j_{\rm coll} = \frac{d-(x_{j+1}-x_j)}{\theta_{j+1}/m(x_{j+1})-\theta_j/m(x_j)}+\mathcal{O}({\rm d}t^2) .
    \label{eq:collisional time}
\end{equation}
Hence, at each time step, we evaluate the minimal collisional time $t_{\rm coll}=\{{\rm min }\, t^j_{\rm coll}, j=1,...,N-1\}$ between particles. As long as $t_{\rm coll}>{\rm d}t$ is valid, we perform the Runge-Kutta time evolution using equations \ref{eq: Runge-Kutta HR} with time step ${\rm d}t$. When $t_{\rm coll}<{\rm d}t$, we perform the Runge-Kutta time evolution with time step $t_{\rm coll}$ and then we perform the collision.
\subsection{Collisions between particles}
In the hard rod gas, the particles experience contact interaction that conserves energy and momentum. Hence, two rods $\{P_j,P_{j+1}\}$ interact only if $x_{j+1}=x_j+d$.\\
Considering the scattering between particles $\{P_j,P_{j+1}\}$ with initial coordinates $\{x_j,\theta_j\}$, $\{x_{j+1}=x_j+d,\theta_{j+1}\}$, their momenta after the collision will be
\begin{equation}
    \theta_{j+1}^{\rm f} = \frac{2m(x_{j+1})\theta_j - m(x_j) \theta_{j+1} + m(x_{j+1})\theta_{j+1} }{m(x_j)+m(x_{j+1})},\quad
    \theta_{j}^{\rm f} = \frac{2m(x_{j})\theta_{j+1} - m(x_{j+1}) \theta_{j} + m(x_{j})\theta_{j} }{m(x_{j})+m(x_{j+1})}.
\end{equation}
Thus, in the case of constant mass $m(x)=m_0$, the particles simply exchange their momenta through collisions.

\section{The GHD equation for a hard rod gas with forces}
In this appendix, we introduce the GHD theory for a hard rod gas in the presence of external forces.
In particular, we consider a one-dimensional gas of hard rods with length $d$ and in the presence of the external trapping potential $V_1(x)$ and with a space-dependent mass $m(x)$ such that the bare energy and momentum of a single particle are defined as
\begin{equation}
    \epsilon(\theta,x) = \frac{\theta^2}{2 m(x)}+V_1(x) \quad\mbox{ , }\quad p(\theta,x)=\theta .
\label{eq: 1-particle nrg and mom}
\end{equation}
In the absence of external potentials, the system is clearly integrable \cite{spohn_large_1991, Boldrighini_1, Percus_HR_1969} and it has been proven that at large scale it is described by GHD \cite{Doyon_HR_2017, doyon_soliton_2017}.
In section \ref{sec: Ghd Euler} we define the dressing operator for the hard rod gas, and we describe the GHD at the Euler scale. In section \ref{sec: Ghd N-S} we define the full GHD equation with viscosities and force terms of the system.

\subsection{The GHD equation at Euler scale}
\label{sec: Ghd Euler}
We consider an ensemble of hard rods moving in one dimension with velocities $v$. The quasiparticles are identified with trajectory tracers, following the centres of particles. Whenever two particles collide, the two tracers associated with them, exchange their position. Thus, they compute an instantaneous displacement of $d$ in space, with $d$ equal to the length of the rods. The momenta $\theta$ of the quasiparticles can be identified with the velocities $v$ of the particles. Hence, we can define a local stationary state via the local density of quasiparticles $\rho(\theta,x,t)$ so that $\rho(\theta,x,t){\rm d}x{\rm d}\theta$ is equal to the number of particles in the phase space interval $[(x,x+{\rm d}x), (\theta,\theta+{\rm d}\theta)]$.\\
Since the spatial displacement of the quasiparticles at each collision is not dependent on their momenta, the scattering kernel (in real space) is particularly simple
\begin{equation}
    T_x(\theta,\theta')=-\frac{d}{2\pi}.
\label{eq: T scattering kernel}
\end{equation}
Given the scattering kernel \ref{eq: T scattering kernel}, we can define the dressing operation for single-particle functions $f(\theta,x)$
\begin{equation}
    f^{\rm dr}(\theta,x) = f(\theta,x) - d\int_{\mathbb{R}}{\rm d}\theta\rho(\theta,x,t)f(\theta,x) .
\label{eq:dressing operation}
\end{equation}
Thus, if we use the dressing operation on the identity we have
\begin{equation}
    1^{\rm dr} = 1-\bar\rho(x,t) d \quad\mbox{ with, }\quad\bar\rho(x,t)\equiv\int_{\mathbb{R}}{\rm d}\theta\rho(\theta,x,t) .
\end{equation}
The effective velocity $v^{\rm eff}(\theta,x)$ of a quasiparticle is defined by \cite{Doyon_lecture_notes}
\begin{equation}
    v^{\rm eff}_{[\rho]}(\theta,x)\equiv\frac{(\partial_{\theta}\epsilon)^{\rm dr}}{(\partial_{\theta}p)^{\rm dr}} ,
    \label{eq: GHD veff}
\end{equation}
and the effective acceleration $a^{\rm eff}(\theta,x)$ is defined by
\begin{equation}
    a^{\rm eff}_{[\rho]}(\theta,x)\equiv\frac{(-\partial_{x}\epsilon)^{\rm dr}}{(\partial_{\theta}p)^{\rm dr}}.
    \label{eq: GHD aeff}
\end{equation}
Using the explicit form of the single-particle energy and momentum \ref{eq: 1-particle nrg and mom} and the dressing operation \ref{eq:dressing operation}, we get the expressions for the effective velocity and acceleration for a hard rod gas
\begin{gather}
    v^{\rm eff}_{[\rho]}(\theta,x,t) = \frac{\theta-d\int_{\mathbb{R}}{\rm d}\theta'\theta'\rho(\theta',x,t)}{m(x)\big(1-\bar\rho(x,t) d\big)},
    \label{eq: GHD HR veff}
    \\
    a^{\rm eff}_{[\rho]}(\theta,x,t) = \mathfrak{f}_1(x)+\mathfrak{f}_2(x)\frac{\frac{\theta^2}{2}-d\int_{\mathbb{R}}{\rm d}\theta'\frac{\theta'^2}{2}\rho(\theta',x,t)}{1-\bar\rho(x,t) d},
    \label{eq: GHD HR aeff}
\end{gather}
where we defined $\mathfrak{f}_{1}=-\partial_x V_1(x)$ and $\mathfrak{f}_{2}=-\partial_x m^{-1}(x)$. The expressions \ref{eq: GHD HR veff} and \ref{eq: GHD HR aeff} are equivalent to the components of the flow vector $\vec{J}^{\rm eff}$ defined in the main text.\\
Thus, the GHD equation at the Euler scale reads
\begin{equation}
    \partial_t \rho(\theta,x,t) + \partial_x\big(v^{\rm eff}_{[\rho]}(\theta,x,t)\rho(\theta,x,t)\big) + \partial_{\theta}\big(a^{\rm eff}_{[\rho]}(\theta,x,t)\rho(\theta,x,t)\big) =0 .
\end{equation}

\subsection{The diffusive GHD equation with forces}
\label{sec: Ghd N-S}
In this section, we take into consideration the second-order terms in the hydrodynamic expansion.
The diffusive GHD equation in the main text can be written as \cite{durnin_inhomogeneous_2021, de_nardis_diffusion_2018}
\begin{equation}
\partial_t \rho  + \partial_x(v^{\rm eff}_{[\rho]}\rho) + \partial_{\theta}(a^{\rm eff}_{[\rho]}\rho)   =     \frac{1}{2} \Big[ \partial_x(\boldsymbol{\mathfrak{D}}_{[\rho]}^{(1,1)} \partial_x \rho ) + \partial_x(\boldsymbol{\mathfrak{D}}_{[\rho]}^{(1,2)} \partial_\theta \rho )+ \partial_\theta(\boldsymbol{\mathfrak{D}}_{[\rho]}^{(2,1)} \partial_x \rho )+ \partial_\theta(\boldsymbol{\mathfrak{D}}_{[\rho]}^{(2,2)} \partial_\theta \rho )   \Big],
\label{eq:GHD forces diff}
\end{equation}
where the diffusion matrices $\boldsymbol{\mathfrak{D}}_{[\rho]}$ is a $2\times2$ matrix of operators defined as
\begin{equation}
    \boldsymbol{\mathfrak{D}}_{[\rho]}(\theta,\theta')= d^2 \Big( \delta_{\theta,\theta'} \int_{\mathbb{R}} {\rm d}\kappa \rho(\kappa,x,t)  \vec{g}(\theta,\kappa)-  \rho(\theta,x,t) \vec{g}(\theta,\theta') \Big),
    \label{eq: diffusion matrix 1}
\end{equation}
with $\delta_{\theta,\theta'}$ Dirac delta.
We also defined the kernels $\vec{g}(\theta,\theta')$ as
\begin{equation}
  \vec{g}(\theta,\theta')= d^2 
   \begin{pmatrix}
       |v^{\rm eff }_{[\rho]}(\theta)-v^{\rm eff }_{[\rho]}(\theta')| 
       & 
       {\rm sgn}\big(v^{\rm eff}_{[\rho]}(\theta) - v^{\rm eff}_{[\rho]}(\theta')\big)\big(a^{\rm eff }_{[\rho]}(\theta)-a^{\rm eff }_{[\rho]}(\theta')\big)
       \\
       {\rm sgn}\big(v^{\rm eff}_{[\rho]}(\theta) - v^{\rm eff}_{[\rho]}(\theta')\big)\big(a^{\rm eff }_{[\rho]}(\theta)-a^{\rm eff }_{[\rho]}(\theta')\big)
       &
       \frac{\big(a^{\rm eff }_{[\rho]}(\theta)-a^{\rm eff }_{[\rho]}(\theta')\big)^2}{|v^{\rm eff }_{[\rho]}(\theta)-v^{\rm eff }_{[\rho]}(\theta')|}
       \\
   \end{pmatrix}
   .
\end{equation}

\subsection{Euler and diffusive time scales}
\label{sec:time scales}
Here we consider a system of hard rods prepared in a homogeneous state with distribution of momenta given by two two Gaussian peaks centered in $\theta=\pm \theta_{\rm Bragg}$ and with a given initial temperature $T_0$. Since rapidities have the same units as momentum, the unit of velocity is $v \sim \theta_{\rm Bragg}/m$. The macroscopic length scale of the system is the parameter $\ell$, defining the potential trap and the space dependent mass. Then, the Euler time scale is given by the macroscopic unit of length and velocity : $t_{\rm Eul}= m\ell/\theta_{\rm Bragg}$.\\
To compute the diffusive time scale we firstly consider the case with $m_0=0$ and $V_0\neq0$. In this case, from GHD equation, we know that the diffusive time scale is inversely proportional to the force $F\sim V_0/\ell$ and to the diffusive kernel $ \boldsymbol{\mathfrak{D}}_{[\rho]}^{(x,x)}\sim d^2 \bar\rho/m$.  Since $[\boldsymbol{\mathfrak{D}}_{[\rho]}^{(x,x)} V_0/\ell]^{-1}=[t^{2}/\ell^{2}]$, then the only time scale that we can define using these quantities, together with the unit of velocity and length, is 
\begin{equation}
t_{\rm Diff} = \frac
{\theta_{\rm Bragg}\ell^2}{d^2 \bar\rho V_0}\,.
\end{equation}
Now we consider the case with $V_0=0$ and $m_0\neq 0$. Again, the diffusive time scale is known to be inversely proportional to the to the force $F\sim \mathfrak{f}_2 \theta^2$ and to the diffusive kernel $ \boldsymbol{\mathfrak{D}}_{[\rho]}^{(\theta,\theta)}\sim d^2 \bar\rho m(x) \mathfrak{f}^2_2 \theta^2_{\rm Bragg} $ . Since $[\boldsymbol{\mathfrak{D}}_{[\rho]}^{(x,\theta)} \mathfrak{f}_2 \theta^2]^{-1}=[t^4/m^2\ell^2]$, the only time scale that we can define using this quantity, together with the unit of velocity and length, is $t_{\rm Diff} = (d^2 \bar\rho \ell m(x)^2\mathfrak{f}_2^3 \theta_{\rm Bragg})^{-1}$. Using that $(m^2(x) \mathfrak{f}^3_2)^{-1}\sim \ell^3 m^2 (m-m_0)^2/m_0^3$ we can express the diffusive time scale in terms of $m_0$
\begin{equation}
t_{\rm Diff} = \frac{\ell^2 m^2(m-m_0)^2}{d^2 \bar\rho\theta_{\rm Bragg}m_0^3 }\,.
\end{equation}

\section{Stationarity of thermal distribution under diffusive GHD }
In this appendix, we show that the thermal distribution is stationary under diffusive GHD. Namely, we prove that the thermal distribution $\rho_{\rm th}(\theta,x)$ is a solution of the equation,
\begin{equation}
    -\partial_x (v^{\rm eff}_{[\rho_{\rm th}]}   \rho_{\rm th } ) - \partial_\theta (a_{[\rho_{\rm th}]}^{\rm eff} \rho_{\rm th })    +     \frac{1}{2} \Big[ \partial_x(\boldsymbol{\mathfrak{D}}_{[\rho_{\rm th}]}^{(1,1)}\partial_x \rho_{\rm th } ) + \partial_x(\boldsymbol{\mathfrak{D}}_{[\rho_{\rm th}]}^{(1,2)} \partial_\theta \rho_{\rm th } )+ \partial_{\theta}(\boldsymbol{\mathfrak{D}}_{[\rho_{\rm th}]}^{(2,1)} \partial_x \rho_{\rm th } )+ \partial_{\theta}(\boldsymbol{\mathfrak{D}}_{[\rho_{\rm th}]}^{(2,2)} \partial_\theta \rho_{\rm th } )   \Big]=0 .
    \label{eq:App1 stationarity}
\end{equation}
In order to do that we introduce some explicit expressions for the derivatives of a thermal distribution.  This latter is defined via the function 
\begin{equation}
    \epsilon_{\rm th}(\theta) = \beta w(\theta;x ) - \mu  + \frac{d}{2 \pi} \int d\theta e^{- \epsilon_{\rm th}(\theta)}, \quad \quad 
    w(\theta, x ) = \frac{\theta^2}{2 m(x)} + V_1(x),
\end{equation}
  as $n_{\rm th}(\theta) = e^{- \epsilon_{\rm th}(\theta)}$ at each point $x$. It satisfies the relations
\begin{equation}
    \begin{split}
    \partial_x n_{\rm th} &= (1-d\bar\rho_{\rm th})\beta\mathfrak{f}_1 n_{\rm th} + \beta\mathfrak{f}_2 n_{\rm th}\bigg(\frac{\theta^2}{2}-d\bar e_{\rm th}\bigg),
    \\
    \partial_{\theta} n_{\rm th} &= -\beta\theta n_{\rm th}/m(x),
    \\
    \partial_x \rho_{\rm th} &= -\frac{d\partial_x\bar\rho_{\rm th}}{(1-d\bar\rho_{\rm th})}\rho_{\rm th} + (1-d\bar\rho_{\rm th})\beta\mathfrak{f}_1 \rho_{\rm th} + \beta\mathfrak{f}_2 \rho_{\rm th}\bigg(\frac{\theta^2}{2}-d\bar e_{\rm th}\bigg),
    \\
    \partial_{\theta} \rho_{\rm th} &= -\beta\theta \rho_{\rm th}/m(x),
    \\
    \partial_x \bar\rho_{\rm th} &=(1-d\bar\rho_{\rm th})^2\Big( \beta \mathfrak{f}_1 \bar\rho_{\rm th} + \beta \mathfrak{f}_2 \bar e_{\rm th} \Big),
    \end{split}
    \label{eq:App1 derivatives}
\end{equation}
with $\bar e_{\rm th} \equiv \int \mbox{d}\theta\, \rho_{\rm th}(\theta)\theta^2/2$ and $\bar u_{\rm th} \equiv \int \mbox{d}\theta\, \theta \rho_{\rm th}(\theta)=0$.\\
Firstly, we derive the explicit expressions for the Euler terms in eq. \ref{eq:App1 stationarity} using the formulas \ref{eq:App1 derivatives}
\begin{equation}
        \begin{split}
        \partial_x(v^{\rm eff}_{[\rho_{\rm th}]}\rho_{\rm th})&=\partial_x(\theta n_{\rm th}/m(x) )=\theta (\partial_x n_{\rm th})/m(x) - \theta \mathfrak{f}_2 n_{\rm th},
        \\
        &= -\mathfrak{f}_1 \partial_{\theta}\rho_{\rm th} + \beta\mathfrak{f}_2 \theta\bigg(\frac{\theta^2}{2}-d\bar e_{\rm th}\bigg)n_{\rm th}/m(x)  - \theta \mathfrak{f}_2 n_{\rm th} 
        .
    \end{split}
\end{equation}
\begin{equation}
    \begin{split}
        \partial_{\theta}( a^{\rm{eff}}_{[\rho_{\rm th}]}\rho_{\rm th} )&=\mathfrak{f}_1\partial_{\theta}\rho_{\rm th} -\beta \mathfrak{f}_2\theta\bigg(\frac{\theta^2}{2}-d\bar e_{\rm th}\bigg)n_{\rm th}/m(x) + \theta\mathfrak{f}_2 n_{\rm th}
        \\
        &=-\partial_x(\rho_{\rm th} v^{\rm{eff}}_{[\rho_{\rm th}]}),
    \end{split}
\end{equation}
hence, the thermal distribution is stationary under Euler GHD. Then we take into consideration the diffusive terms of equation \ref{eq:App1 stationarity}, as follows 
\begin{equation}
\begin{split}
    \boldsymbol{\mathfrak{D}}_{[\rho_{\rm th}]}^{(1,1)}\partial_x \rho_{\rm th} &= \int_{\mathbb{R}} {\rm d}\theta'\bigg[ \rho_{\rm th}(\theta')\vec{g}^{(1,1)}_{\theta,\theta'}\partial_x\rho_{\rm th}(\theta) - \rho_{\rm th}(\theta)\vec{g}^{(1,1)}_{\theta,\theta'}\partial_x\rho_{\rm th}(\theta')\bigg]
    \\
    &=\frac{d^2\beta\mathfrak{f}_2}{m(x)\big(1-d\bar\rho_{\rm th}\big)} \int_{\mathbb{R}} {\rm d}\theta'|\theta-\theta'|\bigg(\frac{\theta^2}{2} - \frac{\theta'^2}{2}\bigg)\rho_{\rm th}(\theta)\rho_{\rm th}(\theta'),
\end{split}
\end{equation}
\begin{equation}
\begin{split}
    \boldsymbol{\mathfrak{D}}_{[\rho_{\rm th}]}^{(1,2)}\partial_\theta \rho_{\rm th} &= \int_{\mathbb{R}} {\rm d}\theta'\bigg[ \rho_{\rm th}(\theta')\vec{g}^{(1,2)}_{\theta,\theta'}\partial_{\theta}\rho_{\rm th}(\theta) - \rho_{\rm th}(\theta)\vec{g}^{(1,2)}_{\theta,\theta'}\partial_{\theta}\rho_{\rm th}(\theta')\bigg]
    \\
    &= -\frac{d^2\beta\mathfrak{f}_2}{m(x)\big(1-d\bar\rho_{\rm th}\big)}\int_{\mathbb{R}} {\rm d}\theta'|\theta-\theta'|\bigg(\frac{\theta^2}{2} - \frac{\theta'^2}{2}\bigg)\rho_{\rm th}(\theta)\rho_{\rm th}(\theta')
    \\
    &= - \boldsymbol{\mathfrak{D}}_{[\rho_{\rm th}]}^{(1,1)}\partial_x \rho_{\rm th},
\end{split}
\end{equation}
\begin{equation}
\begin{split}
    \boldsymbol{\mathfrak{D}}_{[\rho_{\rm th}]}^{(2,1)}\partial_x \rho_{\rm th} &= \int_{\mathbb{R}} {\rm d}\theta'\bigg[ \rho_{\rm th}(\theta')\vec{g}^{(2,1)}_{\theta,\theta'}\partial_{x}\rho_{\rm th}(\theta) - \rho_{\rm th}(\theta)\vec{g}^{(2,1)}_{\theta,\theta'}\partial_{x}\rho_{\rm th}(\theta')\bigg]
    \\
    &= \frac{d^2\beta\mathfrak{f}_2^2}{(1-d\bar\rho_{\rm th})}\int_{\mathbb{R}} {\rm d}\theta'\rm{sgn}(\theta-\theta') \bigg(\frac{\theta^2}{2} - \frac{\theta'^2}{2}\bigg)^2\rho_{\rm th}(\theta)\rho_{\rm th}(\theta'),
\end{split}
\end{equation}
\begin{equation}
\begin{split}
    \boldsymbol{\mathfrak{D}}_{[\rho_{\rm th}]}^{(2,2)}\partial_{\theta} \rho_{\rm th} &= \int_{\mathbb{R}} {\rm d}\theta'\bigg[ \rho_{\rm th}(\theta')\vec{g}^{(2,2)}_{\theta,\theta'}\partial_{\theta}\rho_{\rm th}(\theta) - \rho_{\rm th}(\theta)\vec{g}^{(2,2)}_{\theta,\theta'}\partial_{\theta}\rho_{\rm th}(\theta')\bigg]
    \\
    &=-\frac{d^2\beta\mathfrak{f}_2^2}{(1-d\bar\rho_{\rm th})}\int_{\mathbb{R}} {\rm d}\theta'\rm{sgn}(\theta-\theta') \bigg(\frac{\theta^2}{2} - \frac{\theta'^2}{2}\bigg)^2\rho_{\rm th}(\theta)\rho_{th}(\theta')
    \\
    &=-\boldsymbol{\mathfrak{D}}_{[\rho_{\rm th}]}^{(2,1)}\partial_x \rho_{\rm th},
\end{split}
\end{equation}
where we used the relations \ref{eq:App1 derivatives} and the symmetry of the operators $\boldsymbol{\mathfrak{D}}_{[\rho]}^{(i,j)}$ with $i,j\in\{1,2\}$.\\
Thus, we proved that the thermal distribution is stationary also under the diffusive terms of GHD equation
\begin{equation}
    \partial_x(\boldsymbol{\mathfrak{D}}_{[\rho_{\rm th}]}^{(1,1)}\partial_x \rho_{\rm th} + \boldsymbol{\mathfrak{D}}_{[\rho_{\rm th}]}^{(1,2)}\partial_{\theta} \rho_{\rm th}) + \partial_{\theta}(\boldsymbol{\mathfrak{D}}_{[\rho_{\rm th}]}^{(2,1)}\partial_x \rho_{\rm th} + \boldsymbol{\mathfrak{D}}_{[\rho_{\rm th}]}^{(2,2)}\partial_{\theta} \rho_{\rm th})=0 .
\end{equation}





\section{Kinetic picture of GHD equation}
The aim of this appendix is to give a kinetic interpretation of GHD equation. In section \ref{sec:kin theory Euler}, we derive the Euler GHD equation in the presence of external forces using only kinetic arguments. Next, in section \ref{sec:kin theory Diff}, we give a kinetic interpretation of the diffusive terms of equation \ref{eq:GHD forces diff} as diffusive corrections to ballistic quasiparticle spreading.

\subsection{Kinetic derivation of Euler GHD equation}
\label{sec:kin theory Euler}
In this section, we extend the argument presented in \cite{Boldrighini_1} to derive the Euler GHD equation in the presence of external forces. We consider a system of hard rods of length $d$. The one-particle Hamiltonian of the system is
\begin{equation}
        H(\theta,x) = \frac{\theta^2}{2 m(x)}+V_1(x) ,
\label{eq:Hamiltonian}
\end{equation}
with $V_1(x)$ external trapping potential and $m(x)$ space dependent mass.
The Hamilton equations associated to \ref{eq:Hamiltonian} are 
\begin{equation}
\begin{cases}
    \dot{x} = \frac{\theta}{m(x)}, \\
    \dot{\theta} = \mathfrak{f}_1(x) +  \frac{\theta^2}{2} \  \mathfrak{f}_2(x),
\end{cases}
\label{eq:eq motion}
\end{equation}
where we defined $\mathfrak{f}_{1}\equiv-\partial_x V_{1}(x)$ and $\mathfrak{f}_{2}\equiv-\partial_x m^{-1}(x)$.
Considering the system in the thermodynamic limit, we introduce $\rho(\theta,x,t)$ as the average particle density of the gas, namely $\rho(\theta,x,t){\rm d}x{\rm d}\theta{\rm d}t$ is equal to the number of particles in the interval $[(x,x+{\rm d}x),(\theta,\theta+{\rm d}\theta),(t,t+{\rm d}t)]$. The total particle density is given by \begin{equation}
    \bar\rho(x,t)=\int_{\mathbb{R}}{\rm d}\theta \rho(\theta,x,t).  
\end{equation}
The particles are also labelled to exchange their position during the elastic collisions. Namely, each particle moves freely, according to equations \ref{eq:eq motion}, except for jumps in position and momentum at each collision \cite{Boldrighini_1, Doyon_HR_2017}.
On a timescale much larger than the mean free time, the particles will move with an effective velocity and effective acceleration that depend on all the collisions that they make. In particular, each time a particle collides from the left (right) with another one, its position is shifted to the right (left) by $d$.
Hence, the effective velocity of a particle having momentum $\theta$ must be equal to,
\begin{equation}
    v^{\rm eff}(\theta,x,t) = \frac{\theta}{m(x)}+ d\int_{\mathbb{R}}{\rm d}\theta'\Big(n_+(\theta, \theta')-n_-(\theta, \theta')\Big),
\label{eq:kinetic veff def}
\end{equation}
where $n_{+(-)}(\theta, \theta'){\rm d}t{\rm d}\theta'$ is the probability of a particle with momentum $\theta$ making a scattering with a particle with momentum in $[\theta',\theta'+{\rm d}\theta']$ from right (left) in the time step $[t,t+{\rm d}t]$.
The average distance $\bar d(x,t)$ between particles in the interval $[(x,x+{\rm d}x),(t,t+{\rm d}t)]$ is given by the ratio between the total free space and the total number of particles, 
\begin{equation}
    \bar d(x,t)=\frac{{\rm d}x{\rm d}t- d \bar\rho(x,t) {\rm d}x{\rm d}t}{\bar\rho(x,t){\rm d}x{\rm d}t}=\frac{1-d \bar\rho(x,t)}{\bar\rho(x,t)}.
\end{equation}
The probability of the particle firstly colliding with another one having velocity in the range $[\theta',\theta'+{\rm d}\theta']$ is $\big(\rho(\theta',x,t)/\bar\rho(x,t)\big){\rm d}\theta'$.
Thus, assuming that the space-dependent mass $m(x)$ is slowly varying compared to the dimension $d$ of the rods, the probability of having this scattering in the time interval $[t,t+{\rm d}t]$ is equal to 
\begin{equation}
\begin{split}
    d\Big(n_+(\theta, \theta')-n_-(\theta, \theta')\Big){\rm d}t{\rm d}\theta' &= d\frac{\rho(\theta',x,t)}{\bar\rho(x,t)}{\rm d}\theta'\frac{\big(\theta/m(x)-\theta'/m(x+d))) }{\bar d(x,t)}{\rm d}t
    \\
    &=
    d\frac{\rho(\theta',x,t)}{m(x)\big(1-d \bar\rho(x,t)\big)}(\theta-\theta'){\rm d}t{\rm d}\theta' + \mathcal{O}(d\partial_x m^{-1}(x)).
\end{split}
\label{eq:scattering probability}
\end{equation}
Using the probability \ref{eq:scattering probability} in eq. \ref{eq:kinetic veff def} we get the explicit expression for the effective velocity
\begin{equation}
    v^{\rm eff}(\theta,x,t) = \frac{\theta}{m(x)} + \frac{d}{m(x)\big(1-d \bar\rho(x,t)\big)}\int_{\mathbb{R}}{\rm d}\theta'\rho(\theta',x,t)(\theta-\theta'),
\label{eq:kinetic effective velocity}
\end{equation}
which is equal to the one previously shown in eq. \ref{eq: GHD HR veff}.
The effective acceleration is expected to have a similar form as \ref{eq:kinetic veff def}, but with a different kernel $T_\theta(\theta,\theta')$ defined as the jump of momenta at each scattering, 
\begin{equation}
\begin{split}
    a^{\rm eff}(\theta,x,t)&=
     a^{\rm br}(\theta,x) + \int_{\mathbb{R}}{\rm d}\theta' T_\theta(\theta,\theta')\Big(n_+(\theta, \theta')-n_-(\theta, \theta')\Big)
    \\
    &=a^{\rm br}(\theta,x)+\int_{\mathbb{R}} {\rm d}\theta' T_\theta(\theta,\theta') \frac{\rho(\theta',x,t)}{m(x)\big(1-d \bar\rho(x,t)\big)} (\theta-\theta'),
\end{split}
\label{eq:kinetic effective acceleration}
\end{equation}
where $a^{\rm br}(\theta,x)\equiv \mathfrak{f}_1(x) +  \frac{\theta^2}{2} \mathfrak{f}_2(x)$ is the bare acceleration shown in eq. \ref{eq:eq motion}. We now derive an explicit expression for the kernel $T_\theta(\theta,\theta')$.
Since the collisions are elastic, they conserve energy and momentum of the two particles that are colliding. Thus, if a particle of momentum $\theta$ collides with a particle with $\theta'$, its final momentum will be
\begin{equation}
    \theta^{\rm f} = \frac{2m(x+d)\theta - m(x) \theta' + m(x+d)\theta' }{m(x+d)+m(x)}.
\label{eq:scattering kernel 1}
\end{equation}
Assuming that $m^{-1}(x)$ is slow varying compared to the dimension $d$ of the rods, we can expand it in derivatives: $m^{-1}(x+d)=m^{-1}(x)-d\mathfrak{f}_2(x)+\mathcal{O}\big(d^2\partial_x^2 m^{-1}(x)\big)$. Hence, the series expansion of $m(x+d)$ around $d=0$ is
\begin{equation}
    \frac{m(x+d)}{m(x)}= 1 + d\mathfrak{f}_2(x)m(x) + \mathcal{O}\big(d^2(\partial_x m^{-1})^2/m(x), d^2(\partial^2_x m^{-1})/m(x)\big).
\end{equation}
Expanding up to first order the equation \ref{eq:scattering kernel 1} we get
\begin{equation}
    \theta^{f} = \theta + d \mathfrak{f}_2 m(x) (\theta' + \theta)+ \mathcal{O}\big(d^2(\partial_x m^{-1})^2/m, d^2\partial^2_x m^{-1}/m\big) .
\label{eq:scattering kernel 2}
\end{equation}
Thus, from the linearization \ref{eq:scattering kernel 2} of equation \ref{eq:scattering kernel 1} we can express the scattering kernel in momentum space as
\begin{equation}
    T_\theta(\theta,\theta') \equiv \theta^{f} - \theta \simeq d \mathfrak{f}_2 m(x) \frac{\theta' + \theta}{2} .
\label{eq:scattering kernel 3}
\end{equation}
Finally, using the formula \ref{eq:scattering kernel 3} in the effective acceleration \ref{eq:kinetic effective acceleration} we get 
\begin{equation}
\begin{split}
    a^{\rm eff}(\theta,x,t)
    &=
    a^{\rm br}(\theta,x)+d\mathfrak{f}_2(x)\int_{\mathbb{R}} {\rm d}\theta' \frac{\rho(\theta',x,t)}{1-d \bar\rho(x,t)} \bigg(\frac{\theta^2}{2}-\frac{\theta'^2}{2}\bigg)
    \\
    &=
    \mathfrak{f}_1(x)+\mathfrak{f}_2(x)\frac{\theta^2/2-d\int{\rm d}\theta'\rho(\theta',x,t)\theta'^2/2}{1-d \bar\rho(x,t)},
\end{split}
\label{eq:kinetic effective acceleration 4}
\end{equation}
which is equal to the formula \ref{eq: GHD HR aeff}.

Now we impose the local conservation of mass on a fluid cell which is initially at time $t$ in the interval $[(x,x+{\rm d}x),(\theta,\theta+{\rm d}\theta)]$ \cite{Boldrighini_1}.
After a time step ${\rm d}t$, the initial interval is transformed into $\big[\big(x+v^{\rm eff}(\theta,x,t){\rm d}t \,, x+{\rm d}x+v^{\rm eff}(\theta,x,t){\rm d}t\big),\big(\theta+a^{\rm eff}(\theta,x,t){\rm d}t \,,\theta+{\rm d}\theta+a^{\rm eff}(\theta,x,t){\rm d}t\big)\big]$. 
Imposing the conservation of mass, we find the equation
\begin{multline}
    \rho(\theta,x,t){\rm d}x{\rm d}\theta
    =
    \rho\big(\theta+a^{\rm eff}(\theta,x,t){\rm d}t\,, x+v^{\rm eff}(\theta,x,t){\rm d}t \,,t+{\rm d}t\big)\times 
    \\
    \times \big[ {\rm d}x + v^{\rm{\rm eff}}(x+{\rm d}\theta,x,t){\rm d}t-v^{\rm{\rm eff}}(\theta,x,t){\rm d}t \big] \big[ {\rm d}\theta + a^{\rm eff}(\theta,x+{\rm d}\theta,t){\rm d}t-a^{\rm eff}(\theta,x,t){\rm d}t\big],
\end{multline}
where $v^{\rm eff}$ and $a^{\rm eff}$ are defined in eq. \ref{eq:kinetic effective velocity} and \ref{eq:kinetic effective acceleration 4}.
Taking into consideration only first-order terms we get
\begin{multline}
    \frac{\rho(\theta,x,t+{\rm d}t)-\rho(\theta,x,t)}{{\rm d}t}+v^{\rm eff}(\theta,x,t)\frac{\rho(\theta,x+v^{\rm eff}(\theta,x,t){\rm d}t,t)-\rho(\theta,x,t)}{v^{\rm eff}(\theta,x,t){\rm d}t}+
    \\
    +a^{\rm eff}(\theta,x,t)\frac{\rho(\theta+a^{\rm eff}(\theta,x,t){\rm d}t,x,t)-\rho(\theta,x,t)}{a^{\rm eff}(\theta,x,t){\rm d}t}
    +
    \\
    +\rho(\theta,x,t)\bigg(\frac{v^{\rm eff}(\theta+{\rm d}\theta,x,t)-v^{\rm eff}(\theta,x,t)}{{\rm d}x} 
    + \frac{a^{\rm eff}(\theta+{\rm d}\theta,x,t)-a^{\rm eff}(\theta,x,t)}{{\rm d}\theta}
    \bigg)=0.
\label{eq:Kinetic euler 1}
\end{multline}
Hence, the equation \ref{eq:Kinetic euler 1} is formally identical to the Euler GHD equation
\begin{equation}
    \partial_t\rho(\theta,x,t) + \partial_x\big(v^{\rm eff}(\theta,x,t)\rho(\theta,x,t)\big)+ \partial_{\theta}\big(a^{\rm eff}(\theta,x,t)\rho(\theta,x,t)\big) =0 .
\end{equation}

\subsection{Diffusive corrections to ballistic quasiparticle spreading}
\label{sec:kin theory Diff}
The purpose of this section is to give a kinetic interpretation of the diffusive terms of equation \ref{eq:GHD forces diff} as diffusive corrections to ballistic particle spreading. In particular, we will generalize the argument presented in \cite{Gopalakrishnan_2018} in the case of a system with free particle Hamiltonian \ref{eq:Hamiltonian}.

We consider the trajectory $(x(t), \theta(t))$ of a single particle on the phase space. At the leading order, the particle moves ballistically with $(\dot x(t),\dot\theta(t))\simeq (v^{\rm eff}(\theta,x),a^{\rm eff}(\theta,x))$, where the physical interpretation of these relations is explained in appendix \ref{sec:kin theory Euler}. Since this ballistic spreading depends on the interaction with the other particles in the system, it is subjected to a diffusive broadening due to the fluctuations of the statistical ensemble \cite{Gopalakrishnan_2018}.

Our aim is to compute the variance of the fluctuations $\vec{R}\equiv(\delta x,\delta\theta)$ of the quasiparticle's trajectory in the phase space induced by
a fluctuation in the density of particles at $(\theta',x)$. Namely, we want to compute 
\begin{equation}
    \langle \vec{R}\otimes\vec{R}(\theta,x,t)\rangle^c\big\vert_{\theta'} = \begin{pmatrix}
        \langle\delta x^2(\theta,x,t)\rangle^c\vert_{\theta'} 
        & 
        \langle\delta x\delta\theta(\theta,x,t)\rangle^c\vert_{\theta'}
        \\
        \langle\delta x\delta\theta(\theta,x,t)\rangle^c\vert_{\theta'}
        &
        \langle\delta \theta^2(\theta,x,t)\rangle^c\vert_{\theta'}
    \end{pmatrix} ,
\end{equation}
where we defined $\langle \bullet\rangle^c$ as the connected part of the average over the GGE.
Let us first consider the space variance and express it in terms of density fluctuations
\begin{equation}
    \langle\delta x^2(\theta,x,t)\rangle^c\vert_{\theta'} 
    = 
    t^2\langle(\delta v^{\rm{eff}})^2\rangle^c\vert_{\theta'}  = t^2\bigg(\frac{\delta v^{\rm{eff}}(\theta)}{\delta n(\theta')}\bigg)^2\langle\delta n(\theta')^2\rangle^c,
    \label{eq:x spread Vass 1}
\end{equation}
where $n(\theta,x,t)$ denote the generalized Fermi factor, that for the Hard Rod system is $n(\theta,x,t)= \rho(\theta,x,t)/(1-\bar\rho(x,t) d)$ \cite{Doyon_lecture_notes}.
The density fluctuations of a GGE state, computed over an interval of length $\Lambda$, are diagonal in position and momentum \cite{Gopalakrishnan_2018, Fendley_Cmatrix} and for the hard rod gas are equal to \cite{Doyon_HR_2017}
\begin{equation}
    \langle\delta n(\theta,x)n(\theta',x')\rangle^c \equiv \langle n(\theta,x)n(\theta',x') \rangle - \langle n(\theta,x)\rangle\langle n(\theta',x')  \rangle = \delta(x-x')\delta(\theta-\theta')\frac{n(\theta,x)}{\Lambda\big(1-a\bar\rho(x)\big)}.
\label{eq:correlation matrix}
\end{equation}
Since the collisions between quasiparticles with momenta $\theta$ and $\theta'$ can happen only inside the light-cone defined by the velocity $v^{\rm eff}(\theta)-v^{\rm eff}(\theta')$, the fluctuations must be computed over the region of length $\Lambda=t|v^{\rm eff}(\theta)-v^{\rm eff}(\theta')|$.
The derivative of the effective velocity with respect to the generalized Fermi factor can be computed from equation \ref{eq: GHD veff} and is equal to
\begin{equation}
    \frac{\delta v^{\rm eff}(x, \theta, t)}{\delta n(x, \theta',t)} = d \big(1-\bar\rho(x,t) d\big)\big(v^{\rm eff}(\theta,x,t) - v^{\rm eff}(\theta',x,t)\big).
     \label{eq:derivative veff/n}
\end{equation}

Hence, using the relations \ref{eq:correlation matrix} and \ref{eq:derivative veff/n} inside equation \ref{eq:x spread Vass 1} we get
\begin{equation}
    \langle\delta x^2(\theta,x,t)\rangle^c\vert_{\theta'} = t d^2 \rho(\theta',x,t)|v^{\rm eff}(\theta) - v^{\rm eff}(\theta')|=t\rho(\theta',x,t)\vec{g}_{1,1}(\theta,\theta')
\label{eq:x spread Vass 2}
\end{equation}
where $\vec{g}$ is the matrix defined in the main text. Integrating over the fluctuations of all quasiparticles, we get the full variance of the quasiparticle trajectory,
\begin{equation}
    \langle\delta x^2(\theta,x,t)\rangle^c = t\int_{\mathbb{R}}{\rm d}\theta'\rho(\theta',x,t)\vec{g}_{1,1}(\theta,\theta') = t{\rm diag}(\boldsymbol{\mathfrak{D}}_{[\rho]}^{(1,1)}),
\label{eq: full var space}
\end{equation}
where ${\rm diag}(\boldsymbol{\mathfrak{D}}_{[\rho]}^{(1,1)})$ is the diagonal part of the $(1,1)$ component of the diffusion matrix $\boldsymbol{\mathfrak{D}}_{[\rho]}$.

Now we take into consideration the variance of the trajectories in the momentum space. Using the same argument presented above, we can express it in terms of density fluctuations,
\begin{equation}
    \langle\delta \theta^2(\theta,x,t)\rangle^c\vert_{\theta'} 
    = 
    t^2\langle(\delta a^{\rm{eff}})^2\rangle^c\vert_{\theta'}  = t^2\bigg(\frac{\delta a^{\rm{eff}}(\theta)}{\delta n(\theta')}\bigg)^2\langle\delta n(\theta')^2\rangle^c
    \label{eq:theta spread Vass}
\end{equation}
where the fluctuations $\langle\delta n(\theta')^2\rangle^c$ are evaluated again over an interval of length $\Lambda=|v^{\rm eff}(\theta)-v^{\rm eff}(\theta')|t$.
The derivative of the effective acceleration with respect to the generalized Fermi factor can be computed from equation \ref{eq: GHD aeff} as is
\begin{equation}
    \frac{\delta a^{\rm eff}(\theta,x,t)}{\delta n(\theta',x,t)} = d \big(1-\bar\rho(x,t) d\big)\big(a^{\rm eff}(\theta,x,t) - a^{\rm eff}(\theta',x,t)\big).
\label{eq:derivative aeff/n}
\end{equation}
Using the relations \ref{eq:correlation matrix} and \ref{eq:derivative aeff/n} in eq. \ref{eq:theta spread Vass} we get
\begin{equation}
    \langle\delta \theta^2(\theta,x,t)\rangle^c\vert_{\theta'} 
    = t d^2 \rho(\theta',x,t)\frac{\big(a^{\rm eff}(\theta) - a^{\rm eff}(\theta')\big)^2}{|v^{\rm eff}(\theta) - v^{\rm eff}(\theta')|}=t\rho(\theta',x,t)\vec{g}_{2,2}(\theta,\theta') .
\label{eq: var mom}
\end{equation}
Integrating the quantity computed in eq. \ref{eq: var mom} over all the quasiparticles' fluctuations, we get the full variance of the trajectory in momentum space
\begin{equation}
    \langle\delta \theta^2(\theta,x,t)\rangle^c = t\int_{\mathbb{R}}{\rm d}\theta'\rho(\theta',x,t)\vec{g}_{2,2}(\theta,\theta') = t{\rm diag}(\boldsymbol{\mathfrak{D}}_{[\rho]}^{(2,2)}).
\label{eq: full var mom}
\end{equation}
Finally, we compute the transverse variance $\langle\delta x\delta\theta(\theta,x,t)\rangle^c\vert_{\theta'}$ expressing it in terms of density fluctuations
\begin{equation}
    \langle\delta x\delta\theta(\theta,x,t)\rangle^c\vert_{\theta'} 
    = 
    t^2\langle\delta v^{\rm eff}(\theta)\delta a^{\rm eff}(\theta)\rangle^c\vert_{\theta'} 
    = 
    t^2 \bigg(\frac{\delta v^{\rm eff}(\theta)}{\delta n(\theta')}\frac{\delta a^{\rm eff}(\theta)}{\delta n(\theta')}\bigg)\langle\delta n(\theta')^2\rangle^c .
    \label{eq:x theta spread Vass}
\end{equation}
Using relations \ref{eq:derivative veff/n}, \ref{eq:derivative aeff/n} and \ref{eq:correlation matrix} into equation \ref{eq:x theta spread Vass} we get
\begin{equation}
    \langle\delta x\delta\theta(\theta,x,t)\rangle^c\vert_{\theta'} 
    = 
    t d^2\rho(\theta,x,t) {\rm sgn}\big(v^{\rm eff}(\theta) - v^{\rm eff}(\theta')\big)  \big(a^{\rm eff}(\theta) - a^{\rm eff}(\theta')\big) =
    t\rho(\theta',x,t) \vec{g}_{1,2}(\theta,\theta').
\label{eq: var mom-space}
\end{equation}
Integrating equation \ref{eq: var mom-space} over all possible momenta $\theta'$ we get the full variance of the trajectory in the transverse direction
\begin{equation}
    \langle\delta x\delta \theta(\theta,x,t)\rangle^c = t\int_{\mathbb{R}}{\rm d}\theta'\rho(\theta',x,t)\vec{g}_{1,2}(\theta,\theta') = t{\rm diag}(\boldsymbol{\mathfrak{D}}_{[\rho]}^{(1,2)}).
\label{eq: full var mom-space}
\end{equation}
Thus, from equations \ref{eq: full var space}, \ref{eq: full var mom} and \ref{eq: full var mom-space}, we conclude that the quasiparticles' trajectories broaden diffusively in phase space. Namely, the variance of a single particle's probability distribution in phase space grows linearly in time with a coefficient equal to the diagonal part of the diffusion matrices $\boldsymbol{\mathfrak{D}}_{[\rho]}$ previously defined \ref{eq: diffusion matrix 1}:
\begin{equation}
    \langle \vec{R}\otimes\vec{R}(\theta,x,t)\rangle^c = t \begin{pmatrix}
       {\rm diag}\big(\boldsymbol{\mathfrak{D}}_{[\rho]}^{(1,1)}(\theta,\theta')\big) 
       & 
       {\rm diag}\big(\boldsymbol{\mathfrak{D}}_{[\rho]}^{(1,2)}(\theta,\theta')\big)
       \\
       {\rm diag}\big(\boldsymbol{\mathfrak{D}}_{[\rho]}^{(1,2)}(\theta,\theta')\big)
       &
       {\rm diag}\big(\boldsymbol{\mathfrak{D}}_{[\rho]}^{(2,2)}(\theta,\theta')\big)
    \end{pmatrix}
    .
\end{equation}
\end{appendix}

\end{document}